\documentclass[11pt]{article}
\usepackage{cite}
\usepackage{graphicx}
\usepackage{subfigure}
\usepackage{amsfonts}
\usepackage{amssymb}
\usepackage{amsmath}
\usepackage{heck}
\usepackage{table}
\usepackage{setspace}
\usepackage[all]{xy}
\usepackage{verbatim}
\usepackage{color}
\usepackage{float}
\usepackage{epsfig}
\usepackage{multirow}
\usepackage{cancel}
\usepackage{wrapfig}
\usepackage{wasysym}

\newcommand{\ba}{\begin{eqnarray}}
\newcommand{\ea}{\end{eqnarray}}

\begin{document}

\numberwithin{figure}{section}
\numberwithin{table}{section}

\def\simgt{\mathrel{\lower2.5pt\vbox{\lineskip=0pt\baselineskip=0pt
           \hbox{$>$}\hbox{$\sim$}}}}
\def\simlt{\mathrel{\lower2.5pt\vbox{\lineskip=0pt\baselineskip=0pt
           \hbox{$<$}\hbox{$\sim$}}}}

\begin{titlepage}

\begin{flushright}
\end{flushright}

\begin{center}

{\Large \bf Higgs boson mass, neutrino masses and mixing and keV dark matter 
in an $U(1)_R-$ lepton number model}

\vskip 1cm

{\large Sabyasachi Chakraborty\footnote{Email: tpsc3@iacs.res.in}, 
Sourov Roy\footnote{Email: tpsr@iacs.res.in}}

\vskip 0.5cm

{\it Department of Theoretical Physics, Indian Association
for the Cultivation of Science, \\
2A \& 2B Raja S.C. Mullick Road, Jadavpur, Kolkata-700032, India}

\abstract{
We discuss neutrino masses and mixing in the framework of
a supersymmetric model with an $U(1)_{R}$ symmetry, consisting of
a single right handed neutrino superfield with an appropriate R charge.
The lepton number ($L$) of the standard model fermions are identified
with the negative of their R-charges. As a result, a subset of leptonic
R-parity violating operators can be present and are consistent with the
$U(1)_R$ symmetry. This model can produce one light Dirac neutrino mass at 
the tree level without the need of introducing a very small neutrino Yukawa 
coupling. We analyze the scalar sector of this model in detail paying special
attention to the mass of the lightest Higgs boson. One of the sneutrinos
might acquire a substantial vacuum expectation value leading to interesting
phenomenological consequences. Different sum rules involving the physical
scalar masses are obtained and we show that the lightest Higgs boson mass
receives a contribution proportional to the square of the neutrino 
Yukawa coupling $f$. This allows for a 125 GeV Higgs boson at the tree level 
for $f \sim {\cal O} (1)$ and still having a small tree level mass for the 
active neutrino. In order to fit the experimental results involving neutrino 
masses and mixing angles we introduce a small breaking of $U(1)_R$ symmetry, 
in the context of anomaly mediated supersymmetry breaking. In the presence of 
this small R-symmetry breaking, light neutrino masses receive contributions 
at the one-loop level involving the R-parity violating interactions. We also 
identify the right handed neutrino as a warm dark matter candidate in our 
model. In the case of R-symmetry breaking, the large $f$ case is characterized 
by a few hundred MeV lightest neutralino as an unstable lightest supersymmetric particle (LSP) and we briefly discuss the cosmological implications of such 
a scenario.}

\end{center}
\maketitle

\end{titlepage}
\tableofcontents

\section{Introduction}
\label{intro}
The observation of a new neutral boson, widely believed to be
the first elementary scalar boson of nature, by the CMS and ATLAS
experimental collaborations at the CERN LHC is perhaps the most
important discovery in high energy physics in recent times \cite{atlas,cms}. 
The mass of this particle is measured to be $\sim$ 125 GeV.
Obviously, more data and analysis can confirm whether this is the
Higgs boson of the standard model (SM) or not. On the other hand,
supersymmetric particle searches by ATLAS and CMS for $pp$ collision
at center-of-mass energy, $\sqrt{s}$ = 7 and 8 TeV, has observed no significant
excess over the expected SM background. This has set stringent limits
on the superparticle masses (particularly on the masses of squarks and
gluinos) for a number of supersymmetric models/scenarios 
\cite{cms-susy,atlas-susy}.

At the same time, we have very strong experimental evidences in favor
of neutrino oscillation 
\cite{Schwetz:2011qt,Schwetz:2011zk,GonzalezGarcia:2012sz}. These results 
have firmly established the existence of massive neutrinos and non-trivial 
mixing pattern in the neutrino sector (including the recent discovery 
\cite{Daya-bay, Reno} of a small but non-zero mixing angle $\theta_{13}$). 
Non-vanishing neutrino masses and mixing are very important indications of 
new physics. Naturally, the neutrino sector is a testing ground for various 
models going beyond the SM.

There is also compelling evidence for the existence of dark matter (DM) and 
cosmological observations have measured the relic density of DM with a high 
degree of precision \cite{planck,wmap7}. Nevertheless, the identity of the DM 
remains unknown to date and the potential candidates are, for example, 
the lightest neutralino in an R-parity conserving supersymmetric theory, 
the gravitino, the axino, the axion and the keV sterile neutrino 
\cite{Feng:2010gw}. 

On the theoretical side, supersymmetry (SUSY) is a very popular choice
for new physics. The minimal supersymmetric standard model (MSSM)
with R-parity violation (RPV) is an intrinsically supersymmetric way
of generating observed neutrino masses and mixing pattern. 
There are extensive studies involving MSSM with R-parity violation on neutrino
masses and mixing, under various assumptions, both at the tree and the
loop level \cite{Barbier}.

It is, therefore, tempting to see whether there exist supersymmetric models
which can naturally explain the observed mass of the new scalar boson at
$\sim$ 125 GeV, relax the strong constraints on SUSY particle masses coming 
from the LHC, provide a suitable dark matter candidate and at the same time 
produce neutrino masses and mixing consistent with current data. In this 
direction a class of very interesting models are those with a global 
continuous $U(1)_R$ symmetry \cite{Gregoire, Claudia, Ponton, Kumar, Riva}.
Models with R-symmetry have Dirac gauginos instead of Majorana gauginos and the bounds on the first two generation squarks are somewhat relaxed compared to 
MSSM because of the presence of a Dirac gluino 
\cite{Fayet, Hall, Hall1, Nelson, Chacko, Jack:1999ud, Goodsell, Choi, 
Weiner, Fox, Karim, Benakli:2012cy, Davies, Brust}. 
Flavor and CP violating constraints are also suppressed in this class of models \cite{Kribs}.

Let us mention at this stage that extensive studies have been performed
with Dirac gaugino masses in the R-symmetric limit. In order to have
a Dirac gaugino mass\cite{Karim}, one needs to incorporate a singlet superfield 
$\hat S$ in the adjoint representation of $U(1)_Y$, an $SU(2)_{L}$ triplet 
superfield $\hat T$ (with zero hypercharge), and an $SU(3)_C$ octet 
superfield $\hat O$. The Dirac gaugino masses have also been motivated from 
``supersoft" supersymmetry breaking \cite{Fox-Weiner}. Another notable 
feature of these models are, the absence of trilinear scalar interactions 
({\it A terms}) and also the $\mu$ term, when the R-symmetry is preserved. 
However one can reintroduce these terms by considering the breaking of 
R-symmetry \cite{Goodsell-1}.

Recently, there have been very interesting proposals where the $U(1)_R$ 
symmetry was identified with the lepton number 
\cite{Ponton, Claudia, Gregoire, Kumar, Riva}. A classification 
of phenomenologically interesting R-symmetric models has been performed in 
Ref.\cite{Ponton} showing that leptonic or baryonic RPV operators are allowed 
by such R-symmetries. The role of the down type Higgs is played by the 
sneutrino in these models, which can acquire a significant vacuum expectation 
value (vev) and a light Higgs boson with a mass of $\sim$ 125 GeV can be 
produced \cite{Gregoire, Claudia, Ponton, Riva}. If lepton number is identified 
with the $U(1)_R$ symmetry then even in the presence of leptonic RPV 
operators one cannot generate neutrino Majorana masses violating lepton 
number by two units ($\Delta L$ = 2). One way to avoid such a problem is 
to introduce light Dirac neutrino masses involving gauge singlet neutrino 
superfields with appropriate R-charges.

In this work we take a very interesting and minimalistic approach and 
introduce only a single right handed neutrino superfield in the model. 
We shall discuss in detail at a later stage that this model can produce 
one very light Dirac neutrino \footnote{Although the generic feature of 
this model would be to have a relatively heavy Dirac neutrino, by appropriate 
tuning of some parameters one can have a Dirac neutrino mass less than 
0.1 eV or so.} at the tree level with an Yukawa coupling 
as large as $\sim 10^{-4}$ and in some cases even with an Yukawa coupling 
of $\cal O$(1). In the presence of only a single right-handed neutrino 
the low energy spectrum includes two massless neutrinos and one must think 
of some other mechanism to generate non-zero mass to at least one of these 
massless neutrinos. This can be achieved by introducing a small breaking 
of $U(1)_R$ symmetry. We know that a non-zero gravitino mass
$m_{3/2}$ implies breaking of $U(1)_R$ symmetry. In this work we shall 
consider a small gravitino mass $m_{3/2} \lsim$ 10 GeV in the context of 
anomaly mediated supersymmetry breaking. This ensures that the effects 
of $U(1)_R$ symmetry breaking are also not very large. In fact, the small 
breaking of $R$-symmetry generates small Majorana masses for the gauginos 
as well as trilinear scalar interactions or the $A$-terms\cite{Claudia}. 
We shall show 
in our subsequent analysis that these small $R$-breaking parameters will 
induce non-zero Majorana mass terms for the neutrinos at the tree level as 
well as at the one-loop level. Moreover, gravitino mass in this ballpark 
is also consistent with primordial nucleosynthesis, thermal
leptogenesis and gravitino as a cold dark matter candidate \cite{Ibarra}.

Our analysis shows that in the case of a large neutrino Yukawa coupling $f$, 
an additional tree level contribution to the lightest CP-even Higgs boson 
mass can be obtained, which can be significant for a value of 
$f \sim {\cal O}(1)$. Note that even with such a large value of $f$ 
one can have a small active neutrino mass at the tree level. In the 
presence of this large $f$ the lightest neutralino with a large bino 
component and having a mass of a few hundred MeV becomes the LSP. 
The long-lived gravitino (with a mass $m_{3/2} \sim$10 GeV) is the 
next-to-lightest supersymmetric particle (NLSP) and in order to be 
cosmologically consistent this requires a reheating temperature 
$T_R \lsim 10^6$ GeV.  

One important thing to note is that in this model we can have a sterile
neutrino with mass of the order of a few keV. This can be identified
as a warm dark matter candidate \cite{Abazajian} with appropriate relic
density. We have checked that the active sterile mixing is small and
consistent with the experimental observations of satellite based X-ray
telescopes. Thus we are able to have a situation, where appropriate
values of light neutrino masses and mixing angles are achieved along
with a warm dark matter candidate in the form of sterile neutrino.

The plan of the paper is as follows. First in Section 2, we give an
introduction to the $U(1)_R$ symmetric model with one right handed neutrino
superfield. Section 3 describes the scalar sector of this model along with
the electroweak symmetry breaking conditions. The generation of a $\sim$ 125
GeV Higgs boson mass through one loop radiative corrections is discussed 
in some detail. The neutrino sector in R-symmetric case is discussed in 
Section 4. We introduce R-symmetry breaking in Section 5 in the framework 
of anomaly mediated supersymmetry breaking. Neutrino masses at the tree level 
are discussed in detail accompanied by necessary analytical results. 
In Section 6 we consider the possibility of having a eV scale sterile 
neutrino in this model and discuss its incompatibility to explain the 
LSND\cite{LSND1,LSND2,LSND3} anomaly. Next we present our discussion of a 
keV scale sterile neutrino as a warm dark matter candidate in Section 7. 
Section 8 describes the contribution to neutrino mass matrix at the 
one-loop level in the R-symmetry breaking scenario. We present a 
comprehensive discussion on the results of our numerical analysis of 
neutrino masses and mixing and keV dark matter in Section 9, along with 
the constraints on RPV couplings as a function of the gravitino
mass. The case of large neutrino Yukawa coupling and its relation to the 
tree level Higgs boson mass is discussed in Section 10. We conclude in 
Section 11 along with future outlook. 
\section{$U(1)_{R}$ model with a right handed neutrino}
\label{models}
We consider a minimal extension of the model, introduced in
\cite{Ponton}, with the standard MSSM superfields $\hat H_{u}$,
$\hat H_{d}$, $\hat Q_{i}$, $\hat U_{i}^{c}$, $\hat D_{i}^{c}$,
$\hat L_{i}$, $\hat E_{i}^{c}$ ($i$ = 1, 2, 3), along with one right
handed neutrino superfield $\hat N^{c}$. In addition to this, vector
like $SU(2)_{L}$ doublet superfields $\hat R_{u}$ and $\hat R_{d}$, with
opposite hypercharge ($Y=1, -1$ respectively) are considered. These doublets, 
with appropriate R charge assignments, were originally introduced in order 
to have an anomaly free framework \cite{Kribs} and they are inert in nature. 
The reason for this inertness is, if $\hat R_{u}$ and $\hat R_{d}$ acquires 
a vev, the R-symmetry will be broken spontaneously, which we would like to 
avoid. Again, as mentioned earlier, the bino and wino do not possess Majorana 
masses in the R-symmetry preserving scenario. However, they can acquire 
Dirac masses and for that one has to consider superfields in the adjoint 
representation of the standard model gauge group. These superfields are, 
a singlet $\hat S$, a triplet $\hat T$, under $SU(2)_{L}$ (with zero 
hypercharge), and an octet $\hat O$, under $SU(3)_{c}$. For example, 
by pairing the singlet $\hat S$, with the bino, one obtains a Dirac bino 
mass term and so on.

As discussed in Ref.\cite{Claudia, Ponton}, we identify the R-symmetry with
lepton number symmetry in such a way that the lepton numbers of the SM 
fermions are identified with the negative of their R-charges whereas the 
superpartners of the SM fermions carry lepton numbers same as their R-charges. 
Below we make a table of different superfields in this model with their 
appropriate R-charges.

\vspace{-2mm}
\begin{table}[h!]
\begin{center}
\begin{tabular}{|c|ccccccccccccc|}
\hline
\rule{0mm}{5mm}
& $\hat Q_{i}$ & $\hat U_{i}^{c}$ & $\hat D_{i}^{c}$ & $\hat L_{i}$
& $\hat E_{i}^{c}$ & $\hat H_{u}$ & $\hat H_{d}$ & $\hat R_{u}$
& $\hat R_{d}$ & $\hat S$ & $\hat T$ & $\hat O$ & $\hat N^{c}$
\\[0.3em]
\hline
\rule{0mm}{5mm}
$U(1)_{R}$ & 1 & 1 & 1 & 0 & 2 & 0 & 0 & 2 & 2 & 0 & 0 & 0 & 2 \\
[0.3em]
\hline
\end{tabular}
\end{center}
\vspace{-10pt}
\caption{$U(1)_{R}$ charge assignments to the superfields.}
\label{charges-old}
\vspace{-15pt}
\end{table}
\vspace {5mm}
The superpotential in the R-preserving case becomes,
\ba
W&=&y^{u}_{ij}\hat H_{u}\hat Q_{i}\hat U^{c}_{j}
+\mu_{u}\hat H_{u}\hat R_{d}+f_{i}\hat L_{i}\hat H_{u}\hat N^{c}
+\lambda_{S}\hat S\hat H_{u}\hat R_{d}+2\lambda_{T}
\hat H_{u}\hat T\hat R_{d}-M_{R}\hat N^{c}\hat S+
\mu_{d}\hat R_{u}\hat H_{d}\nonumber \\
&+&\lambda^\prime_{S}\hat S\hat R_{u}\hat H_{d}+\lambda_{ijk}
\hat L_{i}\hat L_{j}\hat E^{c}_{k}
+\lambda^{\prime}_{ijk}\hat L_{i}\hat Q_{j}\hat D^{c}_{k}
+2\lambda^\prime_{T}\hat R_{u}\hat T\hat H_{d}+y^{d}_{ij}\hat H_{d}
\hat Q_{i}\hat D^{c}_{j}+y^{e}_{ij}\hat H_{d}
\hat L_{i}\hat E^{c}_{j} + \lambda_N {\hat N}^c {\hat H}_u {\hat H}_d.
\nonumber \\
\label{superpotential}
\ea
Here the triplet $\hat T$ under $SU(2)_{L}$, is parametrised
as\cite{Espinosa, Mohanty},
\ba
\hat T&=& \sum_{a=1,2,3}\hat T^{(a)},\nonumber \\
&=& \frac{1}{2}
\begin{pmatrix}
\hat T_{0} & \sqrt2 \hat T_{+} \\
\sqrt2 \hat T_{-} & -\hat T_{0}
\end{pmatrix},
\ea
where $\hat T^{(a)}=T_{a}\frac{\sigma^{a}}{2}$, $\sigma^{a}$'s
are the Pauli matrices and we denote $T_{3}=T_{0}$,
$T_{+}=\frac{1}{\sqrt 2}(T_{1}-iT_{2})$ and $T_{-}=\frac{1}{\sqrt 2}
(T_{1}+iT_{2})$. Note that the most general superpotential contributing
to the renormalizable interactions in the Lagrangian includes other terms
such as
\ba
W^\prime = \kappa {\hat N}^c {\hat S} {\hat S} + \eta {\hat N}^c.
\ea
However, in this work for simplicity we will keep $\kappa$ and $\eta$ to be 
equal to zero. It is also important to note that a term of the type 
$\mu^i_L {\hat R}_u {\hat L}_i$ can, in principle, be added to the 
superpotential. Nevertheless, one can rotate away this term using a 
redefinition of the superfields ${\hat L}_i$ and ${\hat H}_d$ such that only a 
linear combination of these superfields couples to ${\hat R}_u$ in the 
superpotential, which we identify as the new ${\hat H}_d$. One must 
remember that the above superpotential (eq.(\ref{superpotential})) is 
written in this rotated basis. 	

\subsection{Soft supersymmetry breaking interactions}
The R-symmetric model discussed above must contain supersymmetry breaking 
in order to make a realistic phenomenological model. In order to do this 
we have to imagine that supersymmetry breaking is not associated with 
R-symmetry breaking in the global supersymmetry case. This can be achieved 
by including both D-term supersymmetry breaking as well as F-term supersymmetry breaking \cite{Ponton, arkani-hamed}. Introducing the
spurion superfields $W^\prime_\alpha = \lambda^\prime_\alpha 
+ \theta_\alpha D^\prime$, the
Dirac gaugino mass terms appear in the Lagrangian as
\ba
{\cal L}^{\rm Dirac}_{\rm gaugino} &=& \int d^2 \theta 
\dfrac{W^\prime_\alpha}{\Lambda}
[\sqrt{2} \kappa_1 ~W_{1 \alpha} {\hat S} + 2\sqrt{2} 
\kappa_2 ~{\rm tr}(W_{2\alpha} {\hat T})
+ 2\sqrt{2} \kappa_3 ~{\rm tr}(W_{3\alpha} {\hat O})] \nonumber \\
&+& h.c.
\label{dirac-gaugino}
\ea
The terms written in eq.(\ref{dirac-gaugino}) preserve a $U(1)_R$ 
symmetry under which the $W_{i\alpha}$ and $W^\prime_{\alpha}$ have R-charge 1. Accordingly, R[$\lambda_{i\alpha}]$ = R[$\lambda^\prime_\alpha$] = 1 
and R[$D^\prime$] = 0.

The integration over the Grassmann coordinates generates the Dirac 
gaugino mass terms as
\ba
{\cal L}^{\rm Dirac}_{\rm gaugino} = -M^D_1 \lambda_1 {\tilde S} 
-M^D_2 \lambda_{2i} {\tilde T}_i
-M^D_3 \lambda_{3a} {\tilde O}_a + ...,
\ea
where $M^D_j = {\kappa_j D^\prime}/\Lambda$ are the Dirac gaugino masses 
with $j = 1, 2, 3$ corresponding to the $U(1)_Y$, $SU(2)_L$, and $SU(3)_C$ 
gauge groups respectively. Here $\Lambda$ is the scale at which SUSY breaking 
is mediated and ${\hat S}$, ${\hat T}$, and ${\hat O}$ are the chiral 
superfields in the adjoint representation of the gauge groups as mentioned 
earlier, with
\ba
{\hat S} &=& S + \sqrt{2}\theta {\tilde S} + ..., \nonumber \\
{\hat T} &=& T + \sqrt{2}\theta {\tilde T} + ..., \nonumber \\
{\hat O} &=& O + \sqrt{2}\theta {\tilde O} + ...
\ea

The $U(1)_R$ conserving soft supersymmetry breaking terms in the scalar sector
are generated by the spurion superfield ${\hat X}$ defined as 
${\hat X} = x + \theta^2 F_X$,
(with $\langle x \rangle = 0$, $\langle F_X \rangle \neq 0$,  
R[${\hat X}$] = 2),
and can be written as
\ba
V_{soft}&=& m^{2}_{H_{u}} H_{u}^{\dagger}H_{u}+m^{2}_{R_{u}}
R_{u}^{\dagger}R_{u}+ m^{2}_{H_{d}}H_{d}^{\dagger}H_{d}
+m^{2}_{R_{d}}R_{d}^{\dagger}R_{d}
+m^{2}_{\tilde L_{i}}\tilde L_{i}^{\dagger}\tilde L_{i}+
m^2_{{\tilde R}_i}{{\tilde l}^\dagger_{Ri} {\tilde l}_{Ri}} +
M_{N}^{2}\tilde N^{c\dagger}\tilde N^{c}\nonumber \\
&+&m_{S}^{2} S^{\dagger}S+2m_{T}^{2} {\rm tr}(T^{\dagger}T)
+2m_O^2 {\rm tr}(O^\dagger O) + (B\mu H_u H_d + {\rm h.c.})
- (b\mu_L^i H_u {\tilde L}_i + {\rm h.c.}) \nonumber \\
&+&(t_{S}S+{\rm h.c.})+\frac{1}{2} b_{S}(S^{2}+{\rm h.c.})
+b_{T} ({\rm tr}(TT) + {\rm h.c.}) +B_O({\rm tr}(OO) + {\rm h.c.}).
\label{soft-scalar-terms}
\ea
We neglect the $U(1)_R$ symmetric scalar trilinear terms in the expression
in eq.(\ref{soft-scalar-terms}) because they are assumed to be suppressed by 
the factor $m_{\rm SUSY}/\Lambda$ where $m_{\rm SUSY} \sim$ 1 TeV and 
$\Lambda$ is typically much larger than $m_{\rm SUSY}$ \cite{Ponton}. It has 
been argued in Ref.\cite{Goodsell-two-loop} that the dangerous $t_S$ parameter 
in scenarios with Dirac gaugino masses are suppressed and that is what we shall consider in the present work, so that this term does not introduce quadratic 
divergence leading to phenomenological disaster. Note that the tadpole 
term $(t_{{\tilde N}^c}{\tilde N}^c +{\rm h.c.})$ is absent from the scalar 
potential because of R-symmetry.

The presence of the bilinear term $b \mu^i_L H_u {\tilde L}^i $ in the scalar 
potential can, in general, lead to non-zero vevs\footnote{In this model 
sneutrino vevs do not violate lepton number and hence they are not 
constrained from the consideration of small Majorana neutrino masses
of active neutrinos.} ($v_i, ~i=1,2,3$) for all the three 
left-handed sneutrinos. However, we can still rotate to a basis in which 
only one of the left-handed sneutrinos acquires a non-zero vev. The rotation 
can be defined as\cite{Barbier}
\ba
\hat L_{i}=\frac{v_{i}}{v_{a}}\hat L_{a} + e_{ib}\hat L_{b}, 
\ea
where ${\hat L}_a$ is the combination of the ${\hat L}_i$ superfields whose 
neutral scalar component gets a non-zero vev $v_a, ~a=1(e)$ whereas the other 
sneutrino fields corresponding to ${\hat L}_b, ~b=2,~3(\mu,~\tau)$ do not 
acquire any vacuum expectation value, that is to say $v_b =0$ for 
$b =2,~3(\mu,~\tau)$. Here $v_a \equiv \sqrt{\sum_i v_i^2}$ and the superfield 
${\hat L}_a$ is defined as 
\ba
{\hat L}_a = \dfrac{1}{v_a}\sum_i v_i {\hat L}_i.
\ea 
The vectors $\{e_{i2}\}$ and $\{e_{i3}\}$ are orthogonal to each other and 
normalized to unity. In addition, they are also orthogonal to the vector 
$\{v_i\}$.

In this basis the term $f_{i}\hat L_{i}\hat H_{u}\hat N^{c}$ in the 
superpotential transforms into $\frac{f_{i}v_{i}}{v_{a}}\hat L_{a}
\hat H_{u}\hat N^{c}+ f_{i}e_{ib}\hat L_{b}\hat H_{u}\hat N^{c}$.
Using the freedom to choose $f_{i}$ such that $f_{i}e_{ib}=0$, the modified 
neutrino Yukawa coupling term in the superpotential looks like 
$f {\hat L}_a {\hat H}_u {\hat N}^c$, where $f \equiv \dfrac{f_i v_i}{v_a}.$ 
Therefore, in this rotated basis the right handed neutrino superfield 
${\hat N}^c$ couples only with ${\hat L}_a, ~a=1(e)$ 
with a coupling strength $f$. Note that in this single sneutrino vev basis 
the soft supersymmetry breaking bilinear term in the scalar potential 
involving the doublet slepton field and the ${\hat H}_u$ field appears as 
$\epsilon^{ij} b\mu_{L}^{a} H_{u}^{i} {\tilde L}_a^j + {\rm h.c.} ~[a=1(e)]$, 
where $\{i,~j\}$ are $SU(2)$ indices with $\epsilon^{12} = -\epsilon^{21} = 1$. The model can be made even more minimal by integrating out the 
fields $\hat R_{u}$ and $\hat H_{d}$, as discussed in \cite{Ponton}. This 
is the situation when the left-handed sneutrino vev $\langle {\tilde \nu}_1 
\rangle$ is much greater\footnote{The constraints on the sneutrino vev can 
be obtained from the precision electroweak measurements of the vector and 
axial-vector coupling of the $Z$ boson to charged leptons as well as from 
the measurements of tau lepton Yukawa coupling \cite{Gregoire,Ponton}.} than 
the down-type Higgs vev $\langle H^0_d \rangle$ and can be achieved with 
$\mu^2_d \gg m^2_{\tilde L}$ where $\mu_d$ is the coefficient of the bilinear 
term $\mu_d {\hat R}_u {\hat H}_d$ in the superpotential and $m^2_{\tilde L}$ 
is the soft mass squared of the left-handed sleptons. In such a case the masses of the charged lepton and down type quarks arise because of the non-zero vev of the left-handed sneutrino.

Furthermore, the trilinear RPV interactions in the superpotential looks like
\ba
\dfrac{1}{2}\lambda_{ijk}\hat L_{i}\hat L_{j}\hat E_{k}^{c} 
+ \lambda^\prime_{ijk}\hat L_{i}\hat Q_{j}\hat D_{k}^{c}&=&
\sum_{b=2,3}\frac{v_i e_{jb}}{v_{a}}\lambda_{ijk} \hat L_{a}\hat L_{b}
\hat E_{k}^{c} 
+ \lambda^\prime_{ijk} \dfrac{v_i}{v_{a}} \hat L_{a}\hat Q_{j}
\hat D_{k}^{c} \nonumber \\
&+& \dfrac{1}{2}(e_{ib}e_{jc}\lambda_{ijk})\hat L_{b}\hat L_{c}\hat E_{k}^{c}
+ \sum_{b=2,3} e_{ib} \lambda^\prime_{ijk} \hat L_{b}\hat Q_{j}\hat D_{k}^{c}. 
\label{rotated-trilinear}
\ea
From eq.(\ref{rotated-trilinear}) we can identify the Yukawa couplings and 
the trilinear R-parity violating couplings in the single sneutrino vev basis as,
\ba
f^l_{bk} = \sum_{ij} \dfrac{v_i e_{jb}}{v_a} \lambda_{ijk}, ~~~~f^d_{jk} 
= \sum_i \dfrac{v_i}{v_a} \lambda^\prime_{ijk},
\ea
\ba
\lambda_{bck} = \sum_{ij} e_{ib}e_{jc}\lambda_{ijk}, ~~~~\lambda^\prime_{bjk} = \sum_i e_{ib} \lambda^\prime_{ijk}.
\ea
In the basis where the charged lepton (${\hat L}_b, ~b=2,3$) and down type 
Yukawa couplings are diagonal, the above superpotential given in 
eq.(\ref{rotated-trilinear}) can be re-written as 
\ba
W^{\rm diag} = \sum_{b=2,3} f^l_b {\hat L}_a {\hat L^\prime}_b 
{\hat E^{\prime c}}_b + \sum_{k=1,2,3} f^d_k {\hat L}_a
{\hat Q^\prime}_k {\hat D^{\prime c}}_k + \sum_{k=1,2,3} \dfrac{1}{2} 
{\tilde \lambda}_{23k}{\hat L^\prime}_2 {\hat L^\prime}_3 
{\hat E^{\prime c}}_k + \sum_{j,k=1,2,3;b=2,3}{\tilde \lambda}^\prime_{bjk}
{\hat L^\prime}_b {\hat Q^\prime}_j {\hat D^{\prime c}}_k.
\ea
Here the prime on the lepton (${\hat L^\prime}_b, ~b=2,3$) and quark 
superfields denotes that they are in the mass-eigenstate basis
\footnote{Note, however, that the mass of the lepton of flavor $a$ 
cannot be generated from the trilinear R-parity violating operators and 
one must invoke R-symmetry preserving supersymmetry breaking 
operators to generate a small mass.} and ${\tilde \lambda}, 
~{\tilde \lambda}^\prime$ are the trilinear R-parity violating 
couplings in that basis. In our subsequent analysis we shall work in this 
mass eigenstate basis and remove the prime from the fields along 
with ${\tilde \lambda}, ~{\tilde \lambda}^\prime \rightarrow \lambda, 
~\lambda^\prime$. Remember that we are also working in a basis where only 
one left-handed sneutrino (corresponding to flavor $a$) gets a vev. 
To reiterate, we observe that these trilinear RPV operators are consistent 
with the R-symmetric superpotential. Nevertheless, this superpotential 
conserves lepton number because of the identification of lepton number with 
R-charges and hence the lepton number violating processes do not constrain 
these trilinear couplings. The flavor structures of these trilinear R-parity 
violating couplings in this model will have important implications in the
context of neutrino masses and other phenomenology as we shall discuss later.

In view of the above discussion it is easy to see that the superpotential and 
the soft SUSY breaking scalar potential include the following terms
\ba
W&=&y_{ij}^{u}\hat H_{u}\hat Q_{i}\hat U_{j}^{c}+\mu_{u}\hat 
H_{u}\hat R_{d}+f\hat L_{a}\hat H_{u}\hat N^{c}+
\lambda_{S}\hat S\hat H_{u}\hat R_{d}+2\lambda_{T}\hat H_{u}
\hat T\hat R_{d}-M_{R}\hat N^{c}\hat S + W^{\rm diag},
\label{final-superpotential}
\ea
and
\ba
V_{soft}&=& m^{2}_{H_{u}}H_{u}^{\dagger}H_{u}+m^{2}_{R_{d}}
R_{d}^{\dagger}R_{d}+m^{2}_{\tilde L_{a}} \tilde L_{a}^{\dagger}
\tilde L_{a}+\sum_{b=2,3} m^{2}_{\tilde L_{b}} \tilde L_{b}^{\dagger}
{\tilde L_{b}}+M_{N}^{2}{\tilde N}^{c\dagger} {\tilde N}^{c}\nonumber \\
&+&m^2_{{\tilde R}_i}{{\tilde l}^\dagger_{Ri} {\tilde l}_{Ri}} +
+m_{S}^{2} S^{\dagger}S+2m_{T}^{2} {\rm tr}(T^{\dagger}T)
+2m_O^2 {\rm tr}(O^\dagger O) 
- (b\mu_L H_u {\tilde L}_a + {\rm h.c.}) \nonumber \\
&+&(t_{S}S+{\rm h.c.})+\frac{1}{2} b_{S}(S^{2}+{\rm h.c.})
+b_{T} ({\rm tr}(TT) + {\rm h.c.}) +B_O({\rm tr}(OO) + {\rm h.c.})
\label{final-softsusy-terms}
\ea

With the above superpotential and the soft SUSY breaking scalar 
potential, in the R-symmetry conserving scenario, we
would now like to analyze the scalar sector of this model consisting of the 
CP-even neutral scalars, CP odd neutral scalars and the charged scalars in 
detail. Here we assume that no CP violating phases exist in the scalar 
potential.

\section{The Scalar Sector}
\label{scalar-sector}
The scalar potential comprises of four different terms.
\ba
V &=& V_{F}+V_{D}+V_{soft}+V_{one-loop},
\ea
where $V_F$ is the $F$-term contribution to the scalar potential, $V_D$ is the
$D$-term contribution, $V_{soft}$ is the soft supersymmetry breaking part and 
$V_{one-loop}$ is the one-loop contribution to the scalar potential. 
The relevant part of the $F$-term contribution is,
\ba
V_{F} &=& \sum_i \left|\frac{\partial W}{\partial \phi_i}\right|^2\nonumber \\
&=&|(\mu_{u}+\lambda_{S}S+\lambda_{T}T_{0})R_{d}^{0}-f \tilde\nu_{L}
\tilde N^{c}+\sqrt 2 \lambda_{T} T_{+}R_{d}^{-}|^2
+|(\mu_{u}+\lambda_{S}S+\lambda_{T}T_{0})H_{u}^{0}-\sqrt 2 \lambda_{T} 
T_{-} H_{u}^{+}|^2\nonumber \\
&+&|\lambda_{S}H_{u}^{0}R_{d}^{0}+M_{R}\tilde N^{c}
-\lambda_{S}H_{u}^{+}R_{d}^{-}|^2
+|\lambda_{T} (H_{u}^{0}R_{d}^{0}+H_{u}^{+}R_{d}^{-})|^2
+|f H_{u}^{0}\tilde N^{c}|^2+|f H_{u}^{+}\tilde N^{c}|^2\nonumber \\
&+&|f (\tilde \nu_{L}H_{u}^{0}-\tilde l_{L}^{-} H_{u}^{+})-M_{R}S|^2
+|(\mu_{u}+\lambda_{S}S-\lambda_{T}T_{0})H_{u}^{+}-
\sqrt 2\lambda_{T}H_{u}^{0}T_{+}|^2
+|\sqrt 2 \lambda_{T} H_{u}^{+}R_{d}^{0}|^2 \nonumber \\
&+&|\sqrt 2\lambda_{T} H_{u}^{0}R_{d}^{-}|^2
+|(\mu_{u}+\lambda_{S}S-\lambda_{T}T_{0})R_{d}^{-}
-f\tilde l_{L}^{-}\tilde N^{c}+\sqrt 2 \lambda_{T}T_{-}R_{d}^{0}|^2
\ea
and the $D$-term contribution is given by 
\ba
\label{gauge}
V_{D}=\frac{1}{2}\sum_{a}D^{a}D^{a}+\frac{1}{2}D_{Y}D_{Y},
\ea
where 
\ba
\label{aux}
D^{a}=g(H_{u}^{\dagger}\tau^{a}H_{u}+\tilde L_{i}^{\dagger}
\tau^{a}\tilde L_{i}+T^{\dagger}\lambda^{a} T)+
\sqrt 2 (M_{2}^{D}T^{a}+M_{2}^{D}T^{a\dagger}).
\ea
The $\tau^{a}$'s are the $SU(2)$ generators in the fundamental representation, 
whereas $\lambda^{a}$'s are the three generators of the $SU(2)$ group in 
adjoint representation. Again, the $D_{Y}$ is computed as,
\ba
\label{aux-1}
D_{Y}=\frac{g^{\prime}}{2}(H_{u}^{+}H_{u}-\tilde L_{i}^{+}\tilde L_{i})
+\sqrt 2 M_{1}^{D}(S+S^{\dagger}).
\ea 
Here $g$ and $g^\prime$ are $SU(2)_L$ and $U(1)_Y$ gauge couplings respectively.

Therefore using eq.(\ref{aux}) and eq.(\ref{aux-1}), we expand 
eq.(\ref{gauge}) and obtain the contribution to the scalar potential from
D-terms as
\ba
V_{D}&=&\frac{g^{\prime 2}}{8}
(|H_{u}^{+}|^2+|H_{u}^{0}|^2-|\tilde \nu_{i}^{0}|^2
-|\tilde l_{i}^{-}|^2)^2
+(M_{1}^{D})^2(S+S^{\dagger})^2
+(M_{2}^{D})^2(T_{0}+T_{0}^{\dagger})^2\nonumber \\
&+&\frac{g^{'}\sqrt 2}{2}M_{1}^{D}(S+S^{\dagger})
(|H_{u}^{+}|^2+|H_{u}^{0}|^2-|\tilde \nu_{i}^{0}|^2
-|\tilde l_{i}^{-}|^2)\nonumber \\
&+&\frac{g^{2}}{8}
(|H_{u}^{+}|^2-|H_{u}^{0}|^2+|\tilde \nu_{i}^{0}|^2
-|\tilde l_{i}^{-}|^2+2|T_{+}|^2-2|T_{-}|^2)^2\nonumber \\
&+&\frac{g^{2}}{8}((H_{u}^{+})^{\ast}H_{u}^{0}+
(\tilde l_{i}^{-})^{\ast}\tilde \nu_{i}^{0}
+\sqrt 2(T_{-}-T_{+})T_{0}^{\ast}+{\rm h.c.})^2
-\frac{(M_{2}^{D})^2}{2}((T_{+}-T_{-})-{\rm h.c.})^2\nonumber \\
&-&\frac{g^{2}}{8}((H_{u}^{0})^{\ast}H_{u}^{+}+
(\tilde l_{i}^{-})^{\ast}\tilde \nu_{i}^{0}+
\sqrt 2T_{0}(T_{+}^{\ast}+T_{-}^{\ast})-{\rm h.c.})^2
+\frac{(M_{2}^{D})^2}{2}((T_{+}+T_{-})+{\rm h.c.})^2\nonumber \\
&+&\frac{g M_{2}^{D}}{2}((T_{+}+T_{-})+{\rm h.c.})((H_{u}^{+})^{\ast}H_{u}^{0}
+(\tilde l_{i}^{-})^{\ast}\tilde \nu_{i}^{0}+
\sqrt 2 (T_{-}-T_{+})T_{0}^{\ast}+{\rm h.c.})\nonumber \\
&-&\frac{g M_{2}^{D}}{2}((T_{+}-T_{-})-{\rm h.c.})
((H_{u}^{0})^{\ast}H_{u}^{+}+(\tilde l_{i}^{-})^{\ast}\tilde \nu_{i}^{0}
+\sqrt 2 T_{0}(T_{+}^{\ast}+T_{-}^{\ast})-{\rm h.c.})\nonumber \\
&+&\frac{\sqrt 2 g M_{2}^{D}}{2}(T_{0}+h.c.)
(|H_{u}^{+}|^2-|H_{u}^{0}|^2+|\tilde \nu_{i}^{0}|^2-|\tilde l_{i}^{-}|^2
+2|T_{+}|^2-2|T_{-}|^2).
\ea

The Soft supersymmetry breaking part of the scalar potential is given by 
eq.(\ref{final-softsusy-terms}) and the dominant radiative corrections to 
the quartic potential are of the form $\frac{1}{2}\delta\lambda_{u} 
(|H_{u}|^2)^2$, $\frac{1}{2}\delta\lambda_{\nu}(|\tilde\nu_{a}|^2)^2$, and 
$\frac{1}{2} \delta\lambda_{3}|H_{u}^{0}|^2|\tilde \nu_{a}|^2$.
The coefficients of these quartic terms are \cite{Karim-1}
\ba
\delta\lambda_{u}&=& \frac{3 y_{t}^{4}}{16\pi^{2}} 
\ln \left(\frac{m_{\tilde t_{1}}m_{\tilde t_{2}}}{m_{t}^2}\right)
+\frac{5\lambda_{T}^{4}}{16\pi^{2}}\ln\left(\frac{m_{T}^{2}}{v^{2}}\right)
+\frac{\lambda_{S}^{4}}{16\pi^{2}}
\ln\left(\frac{m_{S}^2}{v^{2}}\right)\nonumber \\
&-& \frac{1}{16\pi^{2}}\frac{\lambda_{S}^{2}\lambda_{T}^{2}}
{m_{T}^{2}-m_{S}^{2}}(m_{T}^{2}\{\ln\left(\frac{m_{T}^{2}}{v^{2}}\right)-1\}
-m_{S}^{2}\{\ln \left(\frac{m_{S}^{2}}{v^{2}}\right)-1\}),
\ea
\ba
\delta\lambda_{\nu}&=& \frac{3 y_{b}^{4}}{16\pi^{2}} 
\ln \left(\frac{m_{\tilde b_{1}}m_{\tilde b_{2}}}{m_{b}^2}\right)
+\frac{5\lambda_{T}^{4}}{16\pi^{2}}\ln\left(\frac{m_{T}^{2}}{v^{2}}\right)
+\frac{\lambda_{S}^{4}}{16\pi^{2}}
\ln\left(\frac{m_{S}^2}{v^{2}}\right)\nonumber \\
&-& \frac{1}{16\pi^{2}}\frac{\lambda_{S}^{2}\lambda_{T}^{2}}
{m_{T}^{2}-m_{S}^{2}}(m_{T}^{2}\{\ln\left(\frac{m_{T}^{2}}{v^{2}}\right)-1\}
-m_{S}^{2}\{\ln \left(\frac{m_{S}^{2}}{v^{2}}\right)-1\}),
\ea
and finally,
\ba
\delta\lambda_{3}&=& \frac{5 \lambda_{T}^{4}}{32\pi^{2}} 
\ln (\frac{m_{T}^{2}}{v^{2}})
+\frac{1}{32\pi^{2}}\lambda_{S}^{4}
\ln\left(\frac{m_{S}^{2}}{v^{2}}\right)\nonumber \\
&+&\frac{1}{32\pi^{2}}\frac{\lambda_{S}^{2}\lambda_{T}^{2}}{m_{T}^{2}
-m_{S}^{2}}(m_{T}^{2}\{\ln \left(\frac{m_{T}^{2}}{v^{2}}\right)-1\}-
m_{S}^{2}\{\ln\left(\frac{m_{S}^{2}}{v^{2}}\right)-1\}).
\ea
These contributions to the Higgs quartic couplings can be very important 
for the lightest CP-even Higgs boson to have a mass $\sim$ 125 GeV 
for large stop masses and/or large values of the couplings $\lambda_T$ 
and $\lambda_S$. 

\subsection{Symmetry breaking and minimization conditions}
In minimizing the scalar potential we assume that the neutral scalar fields
$H^0_u$, ${\tilde \nu}_a ~(a = 1(e))$, $S$ and $T$ acquire real vacuum 
expectation values $v_u$, $v_a$, $v_s$ and $v_T$, respectively. The scalar 
fields $R_d$ and ${\tilde N}^c$ carry R-charge 2 and they decouple from the 
scalar fields mentioned above carrying R-charge 0. In order to write down 
the minimization conditions, first we split the fields in terms 
of their real and imaginary parts: $H_{u}^{0}=h_{R} +i h_{I}$, 
${\tilde \nu}={\tilde \nu}^a_{R}+i{\tilde \nu}^a_{I}$,
$S=S_{R}+iS_{I}$ and $T=T_{R}+iT_{I}$. The resulting minimization equations 
with respect to $h_{R}$, $\tilde\nu_{R}$, $T_{R}$, and $S_{R}$ fields, are
\ba
&&(m^{2}_{H_{u}}+\mu_{u}^{2})+(b\mu_{L}^{a}-f M_{R}v_{S})(\tan\beta)^{-1}
+\lambda_{S}^{2}v_{S}^{2}+\lambda_{T}^{2}v_{T}^{2}
+2\mu_{u}\lambda_{S}v_{S}+2\mu_{u}\lambda_{T}v_{T}\nonumber \\
&+&2\lambda_{S}\lambda_{T}v_{S}v_{T}+f^{2}v^{2}\cos^{2}\beta
+\sqrt 2(g^{\prime}M_{1}^{D}v_{S}-g M_{2}^{D}v_{T})
+\frac{2\delta\lambda_{u}+\delta\lambda_{3}}{2}v^{2}
\cos^{2} \beta\nonumber \\
&-&\frac{(g^{2}+g^{\prime 2}+4\delta\lambda_{u})}{4}v^{2}\cos 2\beta=0,
\ea
\ba
&&m^{2}_{\tilde L_{a}}+(b\mu_{L}^{a}-f M_{R}v_{S})\tan\beta+
f^{2}v^{2}\sin^{2}\beta
+\frac{g^{2}+g^{\prime 2}-\delta\lambda_{3}
+2\delta\lambda_{\nu}}{4}v^{2}\cos 2\beta\nonumber \\
&+&(\frac{\delta\lambda_{3}+2\delta\lambda_{\nu}}{4})v^{2}
+\sqrt 2(g M_{2}^{D}v_{T}-g^{\prime}M_{1}^{D}v_{S})=0,
\ea
\ba
&&m_{T_{R}}^{2}+\mu_{u}\lambda_{T}\frac{v^{2}}{v_{T}}\sin^{2}\beta
+\lambda^{S}\lambda^{T}\frac{v_{S}}{v_{T}}v^{2}\sin^2\beta
+\lambda_{T}^{2}v^{2}\sin^2\beta
+\frac{g}{\sqrt 2}M_{2}^{D} \frac{v^{2}}{v_{T}}
\cos2\beta=0,
\ea
\ba
&&v_{S}(m^{2}_{S_{R}}+\lambda_{S}^{2}v^{2}\sin^{2}\beta)
+(\mu_{u}\lambda_{S}v^{2}\sin^{2}\beta+\lambda_{S}\lambda_{T}v_{T}
v^{2}\sin^2\beta+t_{S}-\frac{g^{\prime}}
{\sqrt2 }M_{1}^{D}v^{2}\cos2\beta\nonumber \\
&-&\frac{f M_{R}v^{2}\sin2\beta}{2})=0,
\ea
where we identify $m_{T_{R}}^{2}=m_{T}^{2}+b_{T}+4(M_{2}^{D})^{2}$, 
$m_{S_{R}}^{2}=m_{S}^{2}+b_{S}+4(M_{1}^{D})^2 + M^2_R$, 
$\tan\beta = v_u/v_a$ and
$v^2 = v^2_1 + v^2_2$. The W- and the Z-boson masses can be written as
\ba
m^2_W = \dfrac{1}{2}g^2 (v^2 + 4 v_T^2), \nonumber \\
m^2_Z = \dfrac{1}{2} g^2 v^2/\cos^2{\theta_W}. 
\ea
The tree level $\rho$-parameter comes out to be
\ba
\rho \equiv \dfrac{m^2_W}{m^2_Z \cos^2{\theta_W}} = 1 + \dfrac{4v^2_T}{v^2}.
\ea
Electroweak precision measurements of the $\rho$-parameter constrain the 
triplet vev $v_T$ to be $\lsim$ 3 GeV \cite{PDG} and can be taken to be zero 
in the first approximation. 

\subsection{CP-even neutral scalar sector}
\label{CP-even-scalar}
With the help of these minimization equations, it is straightforward 
to write down the neutral CP even scalar squared-mass 
matrix in the basis $(h_{R},\tilde\nu_{R},S_{R},T_{R})$. 
The CP even scalar squared-mass matrix, thus, would be a symmetric 
$4\times 4$ matrix. Note that we are working in the R-symmetry conserving case. 

The elements of the $4\times 4$ CP-even scalar squared-mass matrix 
$M^2_S$ are given by
\ba
\label{CP-even}
(M_S^2)_{11}&=&\frac{(g^{2}+g^{\prime 2})}{2}v^{2}\sin^{2}\beta+
(fM_{R}v_{S}-b\mu_{L}^{a})(\tan\beta)^{-1}
+2\delta\lambda_{u}v^{2}\sin^{2}\beta,\nonumber \\
(M_S^2)_{12}&=&f^{2}v^{2}\sin2\beta+b\mu_{L}^{a}-
\frac{(g^{2}+g^{\prime 2}-2\delta\lambda_{3})}{4}v^{2}
\sin2\beta-fM_{R}v_{S},\nonumber \\
(M_S^2)_{13}&=& 2\lambda_{S}^{2}v_{S}v\sin\beta+2\mu_{u}
\lambda_{S}v\sin\beta+2\lambda_{S}\lambda_{T}v v_{T}\sin\beta
+\sqrt2 g^{\prime}M_{1}^{D}v\sin\beta-fM_{R}v\cos\beta,\nonumber \\
(M_S^2)_{14}&=& 2\lambda_{T}^{2}v_{T}v\sin\beta+2\mu_{u}\lambda_{T}
v\sin\beta+2\lambda_{S}\lambda_{T}v_{S}v\sin\beta-\sqrt 2 gM_{2}^{D}
v\sin\beta,\nonumber \\
(M_S^2)_{22}&=&\frac{(g^{2}+g^{\prime 2})}{2}v^{2}\cos^{2}\beta+
(fM_{R}v_{S}-b\mu_{L}^{a})\tan\beta
+2\delta\lambda_{\nu}v^{2}\cos^{2}\beta,\nonumber \\
(M_S^2)_{23}&=& -\sqrt 2 g^{\prime}M_{1}^{D}v\cos\beta
-fM_{R}v\sin\beta,\nonumber \\
(M_S^2)_{24}&=&\sqrt 2 g M_{2}^{D} v\cos\beta,\nonumber \\
(M_S^2)_{33}&=&-\mu_{u}\lambda_{S}\frac{v^{2}\sin^2\beta}{v_{S}}
-\frac{\lambda_{S}\lambda_{T}v_{T}v^{2}\sin^2\beta}{v_{S}}
-\frac{t_S}{v_{S}}+\frac{g^{\prime}M_{1}^{D}v^{2}\cos2\beta}{\sqrt 2 v_{S}}
+\frac{f M_{R}v^{2}\sin2\beta}{2v_{S}},\nonumber \\
(M_S^2)_{34}&=&\lambda_{S}\lambda_{T}v^{2}\sin^2\beta, \nonumber \\
(M_S^2)_{44}&=&-\mu_{u}\lambda_{T}\frac{v^{2}}{v_{T}}\sin^2\beta
-\lambda_{S}\lambda_{T}v_{S}\frac{v^{2}}{v_{T}}\sin^2\beta
-\frac{g M_{2}^{D}}{\sqrt 2} \frac{v^{2}}{v_{T}}\cos2\beta.
\ea
Since we want to have the lightest CP-even Higgs boson to be doublet-like 
and with a mass around 125 GeV, we would require a small vev $v_S$ of the 
singlet $S$ as well as large radiative corrections to the Higgs boson mass. 
Because of the choices of R-charges of various fields in this model, one 
cannot get tree level contributions to the lightest Higgs boson mass 
proportional to $\lambda_S^2$ and $\lambda_T^2$ as obtained in 
\cite{Karim-1,Goodsell-1}. However, there can be an 
additional contribution to the lightest Higgs boson mass at the tree 
level proportional to the square of the neutrino Yukawa coupling $f$ and 
that can be significant when $f$ is $\cal {O}$(1). We shall discuss 
more on this scenario at a later stage. Note also that the smallness 
of $v_S$ and $v_T$ can be easily obtained by keeping the corresponding 
soft supersymmetry breaking mass terms $m_S$ and $m_T$ somewhat larger 
($\gsim$ a TeV). 
\subsection{CP-odd neutral scalar sector} 
\label{CP-odd}
In the basis $(h_{I},\tilde\nu_{I},S_{I},T_{I})$ the 
elements of the tree-level neutral CP-odd symmetric scalar squared-mass 
matrix $M^2_P$ are
\ba
(M^2_P)_{11}&=&(f M_{R}v_{S}-b\mu_{L}^{a})(\tan\beta)^{-1},\nonumber \\
(M^2_P)_{12}&=&-b\mu_{L}^{a}+fM_{R}v_{S},\nonumber \\
(M^2_P)_{13}&=&-f M_{R}v\cos\beta, \nonumber \\
(M^2_P)_{14}&=&0,\nonumber \\
(M^2_P)_{22}&=&(fM_{R}v_{S}-b\mu_{L}^{a})\tan\beta, \nonumber \\
(M^2_P)_{23}&=& -f M_{R}v\sin\beta, \nonumber \\
(M^2_P)_{24}&=& 0,\nonumber \\
(M^2_P)_{33}&=& \lambda_{S}^{2}v^{2}\sin^2\beta+m_{S_{R}}^2
-2b_{S}-4(M_{1}^{D})^2,\nonumber \\
(M^2_P)_{34}&=& \lambda_{S}\lambda_{T}v^{2}\sin^2\beta, \nonumber \\
(M^2_P)_{44}&=&\lambda_{T}^{2}v^{2}\sin^{2}\beta+m_{T_{R}}^2
-2b_{T}-4(M_{2}^{D})^2.
\ea
The eigenvalues of the CP-odd scalar squared-mass matrix consists of a 
massless Goldstone boson and three physical CP-odd Higgs bosons. Out of these 
three physical Higgs bosons, one is essentially the linear combination 
of $h_{I}$ and $\tilde\nu_{I}$ whereas the other two eigenstates are composed 
mainly of $S_{I}$ and $T_{I}$, the imaginary parts of the singlet $S$ and the 
triplet $T$. 

One can perform the following rotation to separate out the Goldstone mode 
\ba
\begin{pmatrix}
G\\
A\\
S_{I}^{\prime}\\
T_{I}^{\prime}
\end{pmatrix}
=
\begin{pmatrix}
-\sin\beta & \cos\beta &0 &0\\
\cos\beta & \sin\beta &0 &0\\
0 & 0& 1& 0\\
0& 0& 0& 1
\end{pmatrix}
\begin{pmatrix}
h_{I}\\
\tilde\nu_{I}\\
S_{I}\\
T_{I}
\end{pmatrix},
\ea
where $\tan\beta = v_u/v_a$. The $4\times 4$ squared-mass matrix then reduces 
to a $3\times 3$ matrix structure from which one can find out the physical 
CP-odd Higgs bosons. 
\subsection{Charged scalar sector}
\label{Charged-scalar}
In this $U(1)_R$ symmetric case the elements of the tree-level charged 
scalar squared-mass matrix in the basis $(H_{u}^{+}, 
\tilde L_{a}^{-\ast},T^{+},(T^{-})^{\ast})$ are given by ($a = 1(e)$)
\ba
M_{11}^{\pm 2}&=&2\sqrt2 g M_{2}^{D}v_{T}-4v_{S}v_{T}\lambda_{S}\lambda_{T}
-4v_{T}\lambda_{T}\mu_{u}-f^{2}v^{2}\cos^2\beta+\frac{1}{2}g^{2}v^{2}
\cos^{2}\beta\nonumber \\
&+&(-b\mu_{L}^{a}+fM_{R}v_{S})\cot\beta,\nonumber \\
M_{12}^{\pm 2}&=&-b\mu_{L}^{a}+fM_{R}v_{S}-\frac{1}{2}f^{2}v^{2}
\sin 2\beta+\frac{1}{4}g^{2}v^{2}\sin 2\beta, \nonumber \\
M_{13}^{\pm 2}&=&g M_{2}^{D} v\sin\beta-\frac{g^{2}v v_{T}\sin\beta}{\sqrt 2}
-\sqrt 2 v\lambda_{T}(\mu_{u}+v_{S}\lambda_{S}-v_{T}\lambda_{T})
\sin\beta,\nonumber \\
M_{14}^{\pm 2}&=&g M_{2}^{D}v\sin\beta+\frac{g^{2}vv_{T}\sin\beta}{\sqrt 2}
-\sqrt2 v\lambda_{T}(\mu_{u}+v_{S}\lambda_{S}+v_{T}\lambda_{T})
\sin\beta,\nonumber \\
M_{22}^{\pm 2}&=&-2\sqrt2 gM_{2}^{D}v_{T}-f^{2}v^{2}\sin^2\beta
+\frac{1}{2}g^{2}v^{2}\sin^{2}\beta+(-b\mu_{L}^{a}+fM_{R}v_{S})
\tan\beta,\nonumber \\
M_{23}^{\pm 2}&=&g M_{2}^{D}v\cos\beta-\frac{g^{2}vv_{T}\cos\beta}
{\sqrt2},\nonumber \\
M_{24}^{\pm 2}&=&gM_{2}^{D}v\cos\beta+\frac{g^{2}v v_{T}\cos\beta}
{\sqrt 2}, \nonumber \\
M_{33}^{\pm 2}&=&-b_{T}-2(M_{2}^{D})^{2}+g^{2}v_{T}^{2}+
\frac{1}{2}g^{2}v^{2}\cos2\beta-\frac{gM_{2}^{D}v^{2}\cos2\beta}
{\sqrt2 v_{T}}-\frac{v^{2}v_{S}\lambda_{S}\lambda_{T}\sin^{2}\beta}
{v_{T}}\nonumber \\
&+&v^{2}\lambda_{T}^{2}\sin^{2}\beta-\frac{v^{2}
\lambda_{T}\mu_{u}\sin^{2}\beta}{v_{T}},\nonumber \\
M_{34}^{\pm 2}&=&b_{T}+2(M_{2}^{D})^{2}-g^{2}v_{T}^{2},\nonumber \\
M_{44}^{\pm 2}&=&-b_{T}-2(M_{2}^{D})^{2}+g^{2}v_{T}^{2}-
\frac{1}{2}g^{2}v^{2}\cos2\beta-\frac{gM_{2}^{D}v^{2}\cos2\beta}
{\sqrt 2v_{T}}-\frac{v^{2}v_{S}\lambda_{S}\lambda_{T}\sin^{2}\beta}
{v_{T}}\nonumber \\
&-&v^{2}\lambda_{T}^{2}\sin^{2}\beta-\frac{v^{2}\lambda_{T}
\mu_{u}\sin^{2}\beta}{v_{T}}.
\ea
In the limit where the vev of the neutral component of the triplet is very 
small, the triplet essentially decouples from the doublet fields. Considering 
that, the Goldstone mode can be written as
\cite{Espinosa,Goodsell},
\ba
G^{+}&=&(-\sin\beta H_{u}^{+}+\cos\beta \tilde L_{a}^{-\ast}
+a T^{+}+b(T^{-})^{\ast}),
\ea
where a and b represents small admixtures of the triplet fields
with the doublet Higgs-sneutrino block. In order to evaluate the coefficients 
$a$ and $b$, we note that the charged scalar squared-mass matrix follows the 
eigenvalue equation, 
\ba
-M_{11}^{\pm 2}\sin\beta+M_{12}^{\pm 2}\cos\beta+M_{13}^{\pm 2}a
+M_{14}^{\pm 2}b&=&0, \nonumber \\
-M_{12}^{\pm 2}\sin\beta+M_{22}^{\pm 2}\cos\beta+M_{23}^{\pm 2}a
+M_{24}^{\pm 2}b&=&0.
\ea
Solving for $a$ and $b$ in terms of the charged scalar squared-mass matrix 
elements, we find $a=b=\frac{\sqrt 2 v_{T}}{v}$ and finally the expression 
for the Goldstone mode becomes 
\ba
G^{+}&=&\frac{1}{\sqrt\rho}(-\sin\beta H_{u}^{+}+\cos\beta \tilde L_{a}
^{-\ast}+\frac{\sqrt 2 v_{T}}{v}T^{+}
+\frac{\sqrt2 v_{T}}{v}(T^{-})^{\ast}),
\ea
where $\rho$ is the appropriate normalization factor and given by 
$\rho = 1 + \dfrac{4 v^2_T}{v^2}$. The Goldstone boson $G^+$ gives a mass to 
$W^+$ and $G^- \equiv (G^+)^*$ gives a mass to $W^-$. The other states 
orthogonal to $G^+$ are 
\ba
H^{+}&=&\frac{1}{\sqrt\rho}(\cos\beta H_{u}^{+}+\sin\beta \tilde L_{a}
^{-\ast}+\frac{\sqrt2 v_{T}}{v}T^{+}-\frac{\sqrt2 v_{T}}{v}(T^{-})
^{\ast}),\nonumber \\
T_P^{+}&=&\frac{1}{\sqrt\rho}(\frac{\sqrt2 v_{T}}{v}H_{u}^{+}
-\frac{\sqrt2 v_{T}}{v}\tilde L_{a}^{-\ast}+\sin\beta T^{+}
+\cos\beta(T^{-})^{\ast}),\nonumber \\
(T_P^{-})^{\ast}&=&\frac{1}{\sqrt\rho}(\frac{\sqrt2 v_{T}}{v}H_{u}^{+}
+\frac{\sqrt2 v_{T}}{v}\tilde L_{a}^{-\ast}-\cos\beta T^{+}
+\sin\beta(T^{-})^{\ast}).
\ea
Once again we can separate out the Goldstone mode and write down the 
resulting $3\times3$ symmetric charged scalar squared-mass matrix in the 
basis of these orthogonal states (and their charge conjugates) to find out 
the physical charged scalar states.
\subsection{Sum rules}
\label{sum-rules}
We will conclude the discussion on scalar sector by presenting various sum 
rules for this model. Let us look at the CP-even neutral scalar
squared-mass matrix once again and assume that the singlet and triplet vevs are very small. In such a situation these two fields are effectively 
decoupled from the theory and as a result the scalar squared-mass 
matrix becomes, a $2\times 2$ matrix. Under these assumptions, we can write 
down the elements of the neutral CP-even squared-mass matrix 
(for the MSSM case see \cite{Simonsen}) 
in a compact form as (see eq.(\ref{CP-even})),  
\ba
M_{11}^{2}&=&M_{Z}^{2}\sin^2\beta+\xi \cot\beta,\nonumber \\
M_{12}^{2}&=&-\xi+\frac{1}{2}M_{Z}^{2}(\alpha-1)
\sin 2\beta = M_{21}^{2},\nonumber \\
M_{22}^{2}&=&\xi \tan\beta+M_{Z}^{2}\cos^2\beta,
\ea
where we have defined $\alpha=\frac{2f^{2}v^{2}}{M_{Z}^{2}}$ and
$\xi=fM_{R}v_{S}-b\mu_{L}^{a}$. Note that we have kept small terms 
proportional to $v_S$ in this ($2 \times 2$) light CP-even squared-mass matrix. The eigenvalues of this matrix represent the square of the masses of the two 
physical doublet-like Higgs bosons (remember that in this model
the sneutrino of flavor $a$ plays the role of the down type Higgs) and they are given by
\ba
\lambda_\pm=\frac{1}{2}\left[(M_{Z}^{2}+\zeta)\pm \Delta\right],
\ea 
where $\zeta=\frac{2\xi}{\sin2\beta}$ and 
\ba 
\Delta&=&\left[\left(M_{Z}^{2}-\zeta\right)^{2}\cos^2 2\beta+ 
(M_{Z}^{2}(1-\alpha)+\zeta)^{2}\sin^2 2\beta\right]
^{\frac{1}{2}}.
\ea
Similarly, in the decoupling limit of the singlet and triplet fields, 
the CP odd scalar mass matrix has two eigenvalues. One of which 
corresponds to the massless Goldstone boson, whereas the other eigenvalue being
\ba
\zeta&=&\frac{2(-b\mu_{L}^{a}+fM_{R}v_{S})}{\sin 2\beta} \equiv M_{A}^{2}.
\ea

The upper bound on the squared-mass of the lightest CP-even Higgs boson 
($\lambda_- \equiv m^2_h$) will depend on the value of $\Delta$. With the 
help of the inequality \cite{Espinosa}
\ba
\left[a^{2}\cos^{2} 2\beta+b^{2} \sin^{2} 2\beta\right]^
\frac{1}{2}\geqslant \left[a \cos^{2}2\beta+b\sin^{2}2\beta\right],
\ea
we can write down the tree level upper bound on the lightest CP-even Higgs 
boson mass depending on whether $\alpha < 1$ or $\alpha > 1$. 
However, as long as the quantity $\zeta \equiv M^2_A > M_{Z}^{2}$, we find that 
the tree level upper bound on the lightest CP-even Higgs boson mass is 
\ba
m_{h}^{2}\leqslant \left[M_{Z}^{2}\cos^2 2\beta+f^{2}
v^{2}\sin^2 2\beta\right],
\ea
irrespective of whether $\alpha < 1$ or $\alpha > 1$.

It is very interesting to note that the neutrino Yukawa coupling
$f$ provides a tree level correction to the lightest Higgs boson mass. 
We shall discuss later that in our model $f$ can be as large as $\cal O$(1) 
and in that case this large $f$ would certainly provide a significant 
correction to the tree level mass of the lightest Higgs boson, requiring very 
small radiative corrections via the triplet and the singlet as well as from 
the stop loop. 

In a similar way we can obtain a lower bound on the heavy Higgs boson mass 
irrespective of $\alpha$ for $\zeta \equiv M^2_A > M_{Z}^{2}$ and is given by
\ba
m_{H}^{2}\geqslant \left[M_{Z}^{2}\sin^2 2\beta+M^2_A-
f^{2}v^{2}\sin^2 2\beta\right].
\ea
Finally we also obtain a relation between the trace of the CP-even scalar
squared-mass matrix and the trace of the CP-odd scalar squared-mass matrix, 
which differs from that of the MSSM
\ba
{\rm Tr}(M_{S}^{2})&=&{\rm Tr}(M_{P}^{2})+M_{Z}^{2}+2(b_{S}+b_{T})
+4\left[(M_{1}^{D})^{2}+(M_{2}^{D})^2\right].
\ea

Looking at the charged Higgs boson squared-mass matrix in the limit of very 
heavy triplet, we can see that the charged Higgs boson mass ($m_{H^\pm}$) 
can be written in terms of the CP-odd scalar mass ($M_A$) and the W boson 
mass as
\ba
m_{H^{\pm}}^{2}&=&M_{A}^{2}+M_{W}^{2}-f^{2}v^{2}-4v_{T}\lambda_{T}
(\mu_{u}+\lambda_{S}v_{S}).
\ea

Let us also emphasize that we have checked that all the eigenvalues of
the CP-even, CP-odd and charged scalar squared-mass matrices (leaving aside the
Goldstone bosons) come as positive for a minimum. 
\section{Neutrino sector in R-symmetric case}
\label{neutrino-sector}
In the neutral fermion sector we have mixing between the
neutralinos, the active neutrino of flavor $a$, i.e. $\nu_e$ and
the single right-handed neutrino $N^c$ after the electroweak 
symmetry breaking \footnote{In the charged fermion sector the charged lepton 
of flavor $a$ (i.e. $e^\mp$) mixes with the charginos. We shall not discuss
it in this work and refer the reader for a thorough discussion in 
Ref.\cite{Ponton}}. In order to write down the neutral fermion mass matrix, 
we can separate out the relevant part of the Lagrangian as 
$\mathcal{L}=(\psi^{0+})^{T}M_{N}(\psi^{0-})$,
where $\psi^{0+}=(\tilde b^{0},\tilde w^{0},\tilde
R_{d}^{0},N^{c})$, with R-charges $+1$, and $\psi^{0-}
=(\tilde S,\tilde T^{0},\tilde H_{u}^{0},\nu_{e})$, with R-
charges -1. The neutral fermion mass matrix $M_{\chi}^D$ is given by
\ba
M_{\chi}^D=\left(
\begin{array}{cccc}
M_{1}^{D} & 0 & \frac{g^{\prime}v_{u}}{\sqrt 2}& -\frac{g^{\prime}v_{a}}{\sqrt 2}\\
0 & M_{2}^{D} & -\frac{gv_{u}}{\sqrt 2}& \frac{gv_{a}}{\sqrt 2}\\
\lambda_{S}v_{u} & \lambda_{T}v_{u} & \mu_{u}+\lambda_{S}v_{S}
+\lambda_{T}v_{T} & 0\\
M_{R} & 0 & -fv_{a} & -fv_{u}
\end{array} \right).
\label{neutrino-neutralino-matrix}
\ea
The mass matrix $M_{\chi}^D$ can be diagonalized by a biunitary transformation
involving two unitary matrices $V^N$ and $U^N$. This will give rise to
four Dirac mass eigenstates ${\widetilde \chi}^{0+}_i \equiv 
\left( \begin{array}{c} {\widetilde \psi}^{0+}_i \\
\overline {{\widetilde \psi}^{0-}_i} 
\end{array} \right)$, with $i=1,2,3,4$ and ${\widetilde \psi}^{0+}_i = V^N_{ij} \psi^{0+}_j$,
${\widetilde \psi}^{0-}_i = U^N_{ij} \psi^{0-}_j$. The lightest mass eigenstate 
${\widetilde \chi}^{0+}_4$ is identified with the light Dirac neutrino 
eigenstate. The other two active neutrinos remain massless at the tree level in the R-symmetric limit. 
\subsection{Dirac mass of the neutrino}
\label{Dirac-mass}
As we have mentioned, the smallest eigenvalue of the mass matrix $M_\chi^D$ in 
eq.(\ref{neutrino-neutralino-matrix}) corresponds to the light Dirac 
neutrino mass. In order to obtain an analytical expression of this
small mass, we make a series expansion of 
${\rm Det}(M_{\chi}^D- \lambda_{s}\hat I)$, with respect to $\lambda_{s}$ 
and then use the characteristic equation to solve for small 
$\lambda_s$\cite{Morita}
\ba
{\rm Det}(M_{\chi}^D-\lambda_{s}\hat I)&=&{\rm Det}(M_{\chi}^D)-\lambda_{s}
{\rm Det}(M_{\chi}^D){\rm Tr}[(M_{\chi}^D)^{-1}]=0,\nonumber \\
\ea
which implies,
\ba
\label{lambda}
\lambda_{s}&=&\frac{1}{{\rm Tr}[(M_{\chi}^D)^{-1}]}.
\ea
From eq.(\ref{lambda}) we obtain the light Dirac neutrino mass as
\ba
\label{neutrino-mass}
m_{\nu_{e}}^{D}&=&\frac{\left[M_{2}^{D}\gamma \tau+v^{3}
f\sin\beta \omega \right]}
{\left[\gamma(\tau+\sqrt 2 M_{2}^{D}(M_{1}^{D}-fv\sin\beta))
+M_{2}^{D}\tau+(v^{3}f\sin\beta)(g^{\prime}\lambda_{S}-g\lambda_{T})-
v^{2}\sin^2 \beta \omega \right]},
\ea
where,
\ba
\gamma&=&(\mu_{u}+\lambda_{S}v_{S}+\lambda_{T}v_{T}),\nonumber \\
\tau&=&v \cos\beta(g\tan\theta_{W}M_{R}
-\sqrt2 f M_{1}^{D}\tan\beta),\nonumber \\
\omega&=& g(M_{2}^{D}\lambda_{S}\tan\theta_{W}-M_{1}^{D}\lambda_{T}).
\label{gamma-tau-lambda}
\ea
The generic spectrum of this model would include a Dirac neutrino mass
ranging from a few hundred eV to few tens of MeV. However by suitably 
choosing certain relationships involving different parameters, it is 
possible to have a Dirac neutrino mass within 0.1 eV or so.
Therefore, to fit the small neutrino mass, the numerator of 
eq.(\ref{neutrino-mass}) has to be very small. This can be 
achieved by assuming $\omega\rightarrow 0$ and $\tau\rightarrow 0$. 
In this work we shall analyze the case with $\tau=0$ and $\omega\rightarrow 0$. 
However, we shall not discuss the other case $\tau\rightarrow 0$, $\omega=0$ 
in the present work, which can be analyzed in a straightforward way.
The choice $\tau=0$ gives 
\ba
M_{R}=\frac{\sqrt 2f M_{1}^{D}\tan\beta}{g\tan\theta_{W}}.
\label{relation-MR-f}
\ea
Thus eq.(\ref{neutrino-mass}) reduces to (neglecting the term containing 
$\omega$ in the denominator)
\ba
\label{neutrino-mass1}
m_{\nu_{e}}^{D}&=&\frac{\left[v^{3}fg\sin\beta (M_{2}^{D}
\lambda_{S}\tan\theta_{W}-M_{1}^{D}\lambda_{T})\right]}
{\left[\gamma\sqrt 2 M_{2}^{D}(M_{1}^{D}-fV\sin\beta)
+(v^{3}f\sin\beta)(g^{\prime}\lambda_{S}-g\lambda_{T})\right]}.
\ea
The above expression in eq.(\ref{neutrino-mass1}) can be simplified further
by assuming $M_{1}^{D}\gg fv\sin\beta$ and
\ba
\lambda_{T}=\tan\theta_{W}\lambda_{S}.
\label{relation-lamt-lams}
\ea 
With all these alterations in place, the neutrino Dirac mass can be expressed 
in a compact form as
\ba
m_{\nu_{e}}^{D}&=&\frac{v^{3}\sin\beta f g}{\sqrt2 \gamma 
M_{1}^{D} M_{2}^{D}} \lambda_{T}(M_{2}^{D}-M_{1}^{D}).
\ea
As mentioned earlier, that in order to have a small neutrino
mass, the neutrino Yukawa coupling $f$ need not be very small. By 
considering a near degeneracy between the bino and wino Dirac masses
($M_1^D$ and $M_2^D$), it is possible to obtain a small neutrino mass. 
For example, one can choose $f\sim 10^{-5}$, $\lambda_{T}\sim 1$, and 
$(M_{2}^{D}-M_{1}^{D})\sim 10^{-2}$ GeV to accommodate a Dirac neutrino mass 
around $0.1$ eV for $M_1^D$, $M_2^D$ and $\mu_u$ $\sim$ a few hundred GeV.
It is pertinent to mention that the requirement of the degeneracy
of the Dirac gaugino masses is subject to the relations provided in 
eqs.(\ref{relation-MR-f}) and (\ref{relation-lamt-lams}).

However, this near degeneracy between the Dirac gaugino masses can be avoided 
by assuming $\lambda_{T}\sim 10^{-4}$, $f\sim 10^{-4}$. As we have discussed
previously, the triplet coupling $\lambda_{T}$ plays a crucial role to enhance 
Higgs boson mass via radiative corrections. Therefore, in this case of small 
$\lambda_{T}$, we have to consider very heavy stops (with masses around a few 
TeV) for having the lightest CP-even Higgs boson with a mass of 125 GeV. 
In the other case of $\lambda_{T}\sim 1$, the stop masses can be around 
$\sim 700$ GeV and this makes a phenomenologically interesting scenario. On 
the other hand, when $f \sim ~1$, $\lambda_{T}\sim 10^{-6}$ and 
$(M_{2}^{D}-M_{1}^{D})\sim 10^{-2}$ GeV, we can still have a light Dirac 
neutrino mass $\sim$ 0.1 eV and at the same time the lightest Higgs boson 
mass can be $\sim$ 125 GeV at the tree level without requiring multi-TeV stops 
or large triplet coupling for substantial radiative corrections. This is a 
phenomenologically interesting scenario and can be probed further. 

Note that we still have two massless active neutrinos in this model and in 
order to give non-zero masses to these neutrinos one must introduce either 
additional right-handed neutrino superfields with appropriate Yukawa 
interactions or look for R-symmetry breaking effects leading to one-loop
radiative corrections to neutrino masses. Although the first approach is 
interesting and should be explored, in the remaining part of our work we 
shall concentrate on the other approach and introduce a small R-symmetry 
breaking through a non-zero gravitino mass.

\section{R-symmetry breaking}
\label{R-breaking}
So far we have constrained ourselves in the R-symmetry preserving
case. Recent cosmological observations imply a positive but very 
small vacuum energy or cosmological constant associated 
with our universe \cite{planck}. In the context of a spontaneously broken 
supergravity theory in a hidden sector \cite{Martin}, having a very small 
vacuum energy would require a non-zero value of the superpotential in 
vacuum ($<W>$) and that will break R-symmetry because superpotential carry 
non-zero R-charges. Since a non-zero gravitino mass also requires a non-zero 
$<W>$, one can consider the gravitino as the order parameter of the R-symmetry 
breaking. 

The breaking of R-symmetry has to be communicated to the visible sector. 
In this paper, we shall consider the case of anomaly mediated supersymmetry 
breaking playing the role of the messenger of R-symmetry breaking as discussed 
in Ref.\cite{Claudia} and coined as anomaly mediated R-symmetry breaking (AMRB). In this situation, apart from the Majorana gaugino masses and the scalar 
trilinear couplings, all the other R-breaking operators are absent. Finally, 
since we started with an R-symmetry conserving model and afterwards
introduced the breaking of R-symmetry in order to fit neutrino oscillation 
parameters, it is natural to assume that the R-breaking effects are small. This is the case with small gravitino mass as we shall be discussing later in 
more detail. 

In the AMRB scenario, the majorana gaugino masses, generated due to R-breaking, are related to the gravitino mass in the following way
\ba
M_{i}&=&b_{i}\frac{g_{i}^{2}}{16\pi^{2}}m_{3/2},
\ea
where $i=1,2,3$ for bino, wino and gluinos respectively. The 
coefficients are $b_{1}=\frac{33}{5}, b_{2}=1, b_{3}=-3$ and 
one has $g_2 = g$, $g_1 = \sqrt{5/3} g^\prime$.
The third generation trilinear scalar couplings are 
\ba
A_{t}=
\frac{\hat\beta_{h_{t}}}{m_{t}}\frac{m_{3/2}}{16\pi^{2}}v_{u},
A_{b}=
\frac{\hat\beta_{h_{b}}}{m_{b}}\frac{m_{3/2}}{16\pi^{2}}v_{a},
A_{\tau}=
\frac{\hat\beta_{h_{\tau}}}{m_{\tau}}\frac{m_{3/2}}{16\pi^{2}}v_{a},
\label{amsb-a-terms}
\ea
where the $\hat\beta$'s are written in terms of the usual beta 
functions as, $\hat\beta=\frac{\beta}{16\pi^{2}}$ and are given by 
\cite{Gherghetta, Roy},
\ba
\hat\beta_{h_{t}}&=&h_{t}\left(-\frac{13}{15}g_{1}^{2}-3g_{2}^{2}
-\frac{16}{3}g_{3}^{2}+6h_{t}^{2}+h_{b}^{2}\right),\nonumber \\
\hat\beta_{h_{b}}&=&h_{b}\left(-\frac{7}{15}g_{1}^{2}-3g_{2}^{2}
-\frac{16}{3}g_{3}^{2}+h_{t}^{2}+6h_{b}^{2}
+h_{\tau}^{2}\right),\nonumber \\
\hat\beta_{h_{\tau}}&=&h_{\tau}\left(-\frac{9}{5}g_{1}^{2}-3
g_{2}^{2}+3h_{b}^{2}+4h_{\tau}^{2}\right).
\ea
The trilinear scalar couplings for the first two generations can be obtained in 
a straightforward way by replacing the Yukawa couplings appropriately.

So, the Lagrangian containing R-breaking effects\cite{Claudia} in the 
AMRB scenario, can be written as
\ba
\mathcal{L}&=& M_{1}\widetilde b^0 \widetilde b^0 +M_{2}\widetilde 
w^0 \widetilde w^0 +M_{3}\widetilde g\widetilde g +
\sum_{b=2,3} A^l_b {\tilde L}_a {\tilde L}_b {\tilde E^{c}}_b 
+ \sum_{k=1,2,3} A^d_k {\tilde L}_a
{\tilde Q}_k {\tilde D^{c}}_k + \sum_{k=1,2,3} \dfrac{1}{2} 
A^{\lambda}_{23k}{\tilde L}_2 {\tilde L}_3
{\tilde E^{c}}_k \nonumber \\
&+& \sum_{j,k=1,2,3;b=2,3} A^{{\lambda^\prime}}_{bjk}{\tilde L}
_b {\tilde Q}_j {\tilde D^{c}}_k
+A^{\nu}H_{u}\tilde L_{a}\tilde N^{c}
+H_{u} \tilde Q A^u \tilde U^{c}.
\ea
\subsection{Neutralino-neutrino mass matrix in R-breaking scenario}
\label{R-breaking-neutralino}
Our next task is to incorporate the R-breaking effects in 
the neutral fermion mass matrix. Because of the presence of Majorana gaugino 
masses the tree-level neutralino-neutrino mass matrix, written in 
the basis $(\tilde b^{0}, \tilde S, \tilde w^{0},
\tilde T, \tilde R_{d}^{0}, \tilde H_{u}^{0}, N^{c}, \nu_{e})$, is given by
\ba
M_{\chi}^{M}=\left(
\begin{array}{cccccccc}
M_{1} & M_{1}^{D} & 0 & 0 & 0 & \frac{g^{\prime}v_{u}}{\sqrt 2} & 0 
& -\frac{g^{\prime}v_{a}}{\sqrt 2}\\
M_{1}^{D} & 0 & 0 & 0 & \lambda_{S}v_{u} & 0 & M_{R} & 0\\
0 & 0 & M_{2} & M_{2}^{D} & 0 & -\frac{g v_{u}}{\sqrt 2} & 0 & 
\frac{g v_{a}}{\sqrt 2}\\
0 & 0 & M_{2}^{D} & 0 & \lambda_{T}v_{u} & 0 & 0 & 0 \\
0 & \lambda_{S}v_{u} & 0 & \lambda_{T}v_{u} & 0 & \mu_{u}+
\lambda_{S}v_{S}+\lambda_{T}v_{T} & 0 & 0\\
\frac{g^{\prime}v_{u}}{\sqrt 2} & 0 & -\frac{gv_{u}}{\sqrt 2}& 0 &
\mu_{u}+\lambda_{S}v_{S}+\lambda_{T}v_{T} & 0 & -fv_{a} & 0\\
0 & M_{R} & 0 & 0 & 0 & -fv_{a} & 0 & -fv_{u} \\
-\frac{g^{\prime}v_{a}}{\sqrt 2}& 0 & \frac{g v_{a}}{\sqrt 2} & 0 
& 0 & 0 & -fv_{u} & 0
\end{array} \right).
\label{majorana-neutralino}
\ea
In the absence of Majorana gaugino masses ($M_1 = M_2 = 0$), the pure Dirac 
neutrino case discussed in section 4 is recovered from 
eq.(\ref{majorana-neutralino}) and we have one light Dirac neutrino of 
mass $m^D_{\nu_e}$. This is equivalent to saying that we have two 
Majorana neutrinos of mass $-m^D_{\nu_e}$ and $m^D_{\nu_e}$ with opposite CP 
parities\cite{Akhmedov}. 

If the gaugino Majorana mass parameters $M_1$ and $M_2$ are non-zero but 
small compared to the corresponding Dirac gaugino mass parameters $M_1^D$ 
and $M_2^D$ then the pair of light Majorana neutrinos will be quasi-degenerate 
and sometimes called a {\it pseudo-Dirac neutrino}. By increasing the gravitino mass (which means larger $M_1$ and $M_2$) one can generate a larger 
splitting between these two light Majorana neutrino states. Let us discuss 
these two cases in the context of our model, in detail. Note that in the 
absence of $N^c$, the neutralino-neutrino mass matrix cannot produce a non-zero mass of the light neutrino even if the gaugino Majorana mass parameters 
$M_1$ and $M_2$ are non-zero. 
\subsubsection{Case - 1}
In this subsection, we consider a case, where R-breaking effects 
are very small. The two light mass eigenstates of the neutralino-neutrino 
matrix in eq.(\ref{majorana-neutralino}) are almost degenerate, maximally 
mixed and they combine to form a (pseudo)Dirac neutrino. We can evaluate 
the product of these two mass eigenvalues by calculating the ratio of the 
determinants of the full $8\times 8$ matrix and that of the upper 
$6\times 6$ block of $M_\chi^M$, without the $(N^{c},\nu_{e})$ sector.
Assuming small mixing between this neutrino sector with other neutral 
fermions we end up with 
\ba
-\lambda^{2}&=&-\left[\frac{v^{3}\sin\beta f g}{\sqrt 2\gamma
(M_{2}^{D} M^D_1)}\right]^{2} \lambda_T^2 \left(M_{2}^{D}-M_{1}^{D}\right)^{2}.\nonumber \\
&=& -(m^{D}_{\nu_{e}})^{2},
\ea
where $\gamma$ is defined in eq.(\ref{gamma-tau-lambda}) and we have used 
the relations in eq.(\ref{relation-MR-f}) and eq.(\ref{relation-lamt-lams}).
\subsubsection{Case - 2}
Here we are going to consider a relatively lager value
of $m_{3/2}$, which is the order parameter for R-breaking. We 
observe that, with this choice, there is a splitting in masses
of the two light Majorana neutrinos with a relatively smaller mixing
between the two states. The light neutrinos are predominantly right handed 
or left handed and the mass eigenstate $N^{c\prime}$ which is mostly a 
right handed neutrino is heavier than the mass eigenstate 
$\nu_{e}^{\prime}$ with a large left handed component. We shall explicitly 
show this in the section on numerical analysis, but first let us evaluate 
the lightest Majorana neutrino mass, which corresponds to the mass of 
$\nu_{e}^{\prime}$. This can be done by calculating the ratio of the 
determinant of the $8\times 8$ neutralino-neutrino mass matrix 
$M^M_\chi$, to that of the $7\times 7$ upper block of $M^M_\chi$. For a very
small neutrino mass we can assume that the eigenvalues of the $7\times 7$ 
matrix remain unchanged from the seven heavier eigenvalues of the 
$8\times 8$ matrix. This approximation can be safely implemented as long as 
$M_{1}\gg \frac{g^{\prime 2} v_{a}^{2}}{2 M_{1}^{D}}$ and $M_{2}\gg 
\frac{g^{2} v_{a}^{2}}{2 M_{2}^{D}}$. We shall choose the mass of the 
gravitino in such a way that these conditions are satisfied. Therefore the 
light active Majorana neutrino mass at the tree level, in the R-symmetry 
breaking scenario is
\ba
\label{majorana-mass-tree-level}
(m_\nu)_{\rm Tree} &=&-v^{2}\frac{\left[g \lambda_{T} v^{2}(M_{2}^{D}-M_{1}^{D})
\sin\beta\right]^{2}}{\left[M_{1}\alpha^{2}+M_{2}\delta^{2}\right]},
\ea
where
\ba
\alpha&=&\frac{2 M_{1}^{D} M_{2}^{D}\gamma\tan\beta}{g\tan\theta_{w}}
+\sqrt 2  v^{2}\lambda_{S}\tan\beta(M_{1}^{D}\sin^{2}\beta+
M_{2}^{D}\cos^{2}\beta),\nonumber \\
\delta&=&\sqrt2  M_{1}^{D}v^{2}\lambda_{T}\tan\beta.
\ea
In order to obtain eq.(\ref{majorana-mass-tree-level}) we have used once again 
the relations in eq.(\ref{relation-MR-f}) and eq.(\ref{relation-lamt-lams}) and 
$\gamma$ has been defined previously. We can see from eq.
(\ref{majorana-mass-tree-level}) that $m_\nu = 0$ (at the tree level) when 
$M_2^D = M_1^D$ and a small splitting of these Dirac gaugino mass parameters 
will result in a value of $m_\nu$ in the right ballpark provided 
$M_1^D,~M_2^D$ are of the order of a few hundred GeV or 1 TeV with the
couplings $\lambda_T, \lambda_S \sim 10^{-4}$ or so. It is very interesting 
to note that $(m_\nu)_{Tree}$ is independent of the neutrino Yukawa 
coupling $f$. This is an artifact of the relation we have used in 
eq.(\ref{relation-MR-f}) and thus even for a large $f$ $\sim$ ${\cal O}(1)$, 
the tree level Majorana mass of the active neutrino can be kept very small 
with the above choices of parameters. Our approximate analytical result matches very well with the full numerical analysis as described later in this work.

To derive an expression for the mass of the sterile neutrino, 
we work in the region of parameter space, where the active neutrino 
becomes a pure left handed neutrino state. Thus by excluding this left 
handed neutrino state, we are left with a $7\times 7$ neutralino mass 
matrix and the lightest eigenvalue then would correspond to the mass of 
the sterile neutrino. In the limit of large $M_{1}^{D}$, together with 
small couplings \footnote{This particular choice of the 
couplings matches with the benchmark points with heavy stops, considered 
later.} $\lambda_{S}$, $\lambda_{T}$, f and considering only the dominant 
contributions, we eventually obtain the sterile neutrino mass as
\ba
M_{N}^{R}\simeq \left(\frac{M_{1}}{M_{1}^{D}}\right)
\left(\frac{M_{R}}{M_{1}^{D}}\right) M_{R}.
\label{Sterile-mass-1}
\ea

Substituting the expression of $M_{R}$, given in eq.(\ref{relation-MR-f}),
we reduce the sterile neutrino mass in the form,
\ba
M_{N}^{R}= M_{1} \frac{2 f^{2}\tan^{2}\beta}{g^{'2}},
\label{Sterile-mass}
\ea 
which is independent of $M_{1}^{D}$.
\section{eV scale sterile neutrino}
\label{LSND}
The right handed sterile neutrino, introduced in our model can be 
at the eV scale or at the keV scale depending on the relevant model
parameters. In this section we shall analyze the situation when the 
sterile neutrino is considered to have a mass around 1.2 eV. A mass 
of the sterile neutrino, in this range, could in principle 
explain the LSND anomaly. We have discussed in the previous section
that there are two different cases, one where the active and sterile 
neutrinos mix maximally to form a (pseudo)Dirac neutrino, 
and in the other case there is a relatively large mass splitting between the
sterile neutrino and the active neutrino with a very small mixing. In the 
latter situation there are two distinct Majorana neutrinos in the spectrum.
Let us now discuss these two cases separately in the light of the LSND anomaly
\cite{LSND1,LSND2,LSND3}.

\subsection{Pseudo-Dirac case}
When the R-breaking effects are small, the light 
neutrinos are almost degenerate in mass at the tree level 
and with near maximal mixing between the two states. In this case, 
taking into account the possible loop contributions for the 
active neutrinos as well as the sterile neutrino, the neutrino mass matrix 
has a two texture zero structure in the basis $(N_{R}^{\prime},\nu_{e}^{\prime},\nu_{\mu},\nu_{\tau})$, where the prime signifies that these two states combine to form a (pseudo)Dirac neutrino
\ba
\label{texture-1}
\left(
\begin{array}{cccc}
\times&\star&0&0\\
\star &\times &\times &\times\\
0&\times&\times&\times\\
0&\times&\times&\times
\end{array}
\right).
\ea
The asterisks in the (12) and (21) elements symbolise the Dirac
neutrino mass obtained at tree level from the neutralino-neutrino
mass matrix $M^M_\chi$. The crosses in the mass matrix signify the 
contributions to neutrino masses via loop corrections which we shall discuss 
elaborately in the next section. The right handed sterile neutrino mixes 
maximally with the active neutrino in the pseudo-Dirac case. As a result, 
we took into consideration a small mass of the right handed neutrino, 
generated by loops. Finally we have a texture two zero structure of the 
neutrino mass matrix, in the 3+1 scenario\footnote{A detailed study of two 
texture zero neutrino mass matrix structure has been performed in 
\cite{Ghosh:2012pw}}. 

In order to check whether such a texture of neutrino mass matrix is ruled 
out or not, we consider a general neutrino mass matrix in the basis 
$(N^{'}_{R},\nu^{'}_{e},\nu_{\mu},\nu_{\tau})$
\ba
M_{\nu}&=&\left(
\begin{array}{cccc}
M_{ss}&M_{se}&M_{s\mu}&M_{s\tau}\\
M_{es}&M_{ee}&M_{e\mu}&M_{e\tau}\\
M_{\mu s}&M_{\mu e}&M_{\mu\mu}&M_{\mu\tau}\\
M_{\tau s}&M_{\tau e}& M_{\tau\mu}&M_{\tau\tau}
\end{array}
\right).
\ea
This mass matrix can be diagonalised by a $4\times 4$ PMNS matrix $U$
which can be constructed with 6 orthogonal rotation matrices.
For simplicity, let us consider the scenario with no CP violating phases.
The neutrino mass matrix can be obtained from
\ba
\label{Neutrino-mass-matrix-LSND}
M_{\nu}&=&
\left(
\begin{array}{cccc}
U_{s1}&U_{s2}&U_{s3}&U_{s4}\\
U_{e1}&U_{e2}&U_{e3}&U_{e4}\\
U_{\mu 1}&U_{\mu 2}&U_{\mu 3}&U_{\mu 4}\\
U_{\tau 1}&U_{\tau 2}&U_{\tau 3}& U_{\tau 4}\\
\end{array}
\right)
\left(
\begin{array}{cccc}
m_{1}&0&0&0\\
0&m_{2}&0&0\\
0&0&m_{3}&0\\
0&0&0&m_{4}
\end{array}
\right)
\left(
\begin{array}{cccc}
U_{s1}&U_{e1}&U_{\mu 1}&U_{\tau 1}\\
U_{s2}&U_{e2}&U_{\mu 2}&U_{\tau 2}\\
U_{s3}&U_{e3}&U_{\mu 3}&U_{\tau 3}\\
U_{s4}&U_{e4}&U_{\mu 4}&U_{\tau 4}
\end{array}
\right),
\ea
where $m_1, ~m_2, ~m_3, ~m_4$ are the physical neutrino masses and 
$m_1 \gg m_2,m_3,m_4$. We compare this with the two texture zero structure 
given in eq.(\ref{texture-1})and obtain two equations corresponding 
to the zeros in the mass matrix. They are as follows
\ba
\label{mass-terms}
M_{s\mu}&=&m_{1}U_{s1}U_{\mu 1}+m_{2}U_{s2}U_{\mu 2}+m_{3}
U_{s3}U_{\mu 3}+m_{4}U_{s4}U_{\mu 4}=0,\nonumber \\
M_{s\tau}&=&m_{1}U_{s1}U_{\tau 1}+m_{2}U_{s2}U_{\tau 2}+m_{3}
U_{s3}U_{\tau 3}+m_{4}U_{s4}U_{\tau 4}=0.
\ea
These equations can be further simplified with the assumption
that the lightest neutrino mass $m_4$ could be zero. This choice 
is justified as the oscillation experiments are sensitive to the 
mass squared differences. With this simplification, eq.(\ref{mass-terms}) 
reduces to
\ba
\label{modified-mass-terms}
m_{1}U_{s1}U_{\mu 1}+m_{2}U_{s2}U_{\mu 2}+m_{3}U_{s3}U_{\mu 3}&=&0,\nonumber \\
m_{1}U_{s1}U_{\tau 1}+m_{2}U_{s2}U_{\tau 2}+m_{3}U_{s3}U_{\tau 3}&=&0.
\ea 
We notice that eq.(\ref{modified-mass-terms}) contains $m_{1}U_{s1}$, which 
is much larger than all the other terms. Thus no cancellation between this term and the rest can satisfy eq.(\ref{modified-mass-terms}) and we conclude that 
this texture is not viable to explain LSND anomaly. 
\subsection{Majorana case}
\label{Majorana}
In this section we shall consider a different texture 
of the light neutrino mass matrix, where we have one
Majorana neutrino with a tree level mass $\sim 1.2$ eV, but is
composed mainly of the right handed sterile neutrino. The other
Majorana neutrino has a very small mass at the tree level and it is
essentially an active neutrino. Again taking into account possible
loop contributions to the active neutrinos, the three texture zero 
structure\footnote{A detailed study of three texture zero neutrino mass matrix 
structure has been performed in \cite{Zhang, Mona}.} 
of the neutrino mass matrix in the basis 
$(N_{R}^{\prime},\nu_{e}^{\prime},\nu_{\mu},\nu_{\tau})$ is given by
\ba
\left(
\begin{array}{cccc}
\star&0&0&0\\
0&\star&\times&\times\\
0&\times&\times&\times\\
0&\times&\times&\times
\end{array}
\right).
\label{majorana-lsnd}
\ea
The asterisks in the (11) and (22) elements represent the tree level Majorana
masses of $N_{R}^{\prime}$ and $\nu_{e}^{\prime}$ (with additional loop 
contribution in the (22) element) whereas all the other masses 
are generated at the one-loop level. The state $N_{R}^{\prime}$ is mostly
a right handed sterile neutrino and the active sterile mixing in this case is 
negligible.

Comparing the neutrino mass matrix obtained in 
eq.(\ref {Neutrino-mass-matrix-LSND}), with the three texture zero structure of eq.(\ref {majorana-lsnd}), we find
\ba
\label{mass-terms-majorana}
m_{1}U_{s1}U_{i1}+m_{2}U_{s2}U_{i2}+m_{3}U_{s3}U_{i3}
+m_{4}U_{s4}U_{i4}&=&0 ~~~~~~~~~~(i = e, \mu, \tau).
\ea
Again with the assumption of the lightest neutrino mass, $m_{4}=0$, this 
expression can be simplified further. However, as argued in the previous 
section, eq.(\ref {mass-terms-majorana}) cannot be solved by taking
into consideration the neutrino oscillation parameters which satisfy
the LSND anomaly. 

Thus we see that this model as it is, cannot solve the LSND anomaly. 
Nevertheless, in the next section we shall see that by appropriate choice of 
parameters we can fit the three flavor global neutrino data in this model and 
at the same time the sterile neutrino can be accommodated as a keV warm dark 
matter candidate. 
\section{Right handed neutrino as a keV warm dark matter}
\label{warm-dark-matter}
We are considering a situation where the Majorana sterile neutrino 
acquires a tree level mass of the order of a few keV. We work in a 
specific region of parameter space, where R-breaking effects are not 
so large implying that the gravitino mass is around a few GeV 
($m_{3/2}\sim 10$ GeV). There has been a lot of work on model building aspects
of keV sterile neutrino dark matter. For example, keV sterile neutrino dark 
matter has been discussed in gauge extensions of the SM 
\cite{Lindner, Nemevsek:2012cd}, models of composite Dirac neutrinos 
\cite{Grossman:2010iq, Robinson:2012wu}, ${\textrm 331}$ models
\cite{Ky:2005yq, Dinh:2006ia}, models involving Froggatt-Nielsen mechanism 
\cite{Merle:2011yv} and in several other contexts. A review of different 
models/mass generation mechanisms can be found in \cite{Merle:2013gea}. Various other issues related to keV sterile neutrinos can be 
found in \cite{Kusenko:2009up, Abazajian:2012ys}.

Let us give an outline of the case we have considered. The neutrino mass matrix 
in the basis $(N_{R}^{\prime},\nu_{e}^{\prime} ,\nu_{\mu},\nu_{\tau})$ looks 
like
\ba
\left(
\begin{array}{cccc}
\star&0&0&0\\
0&\star&\times&\times\\
0&\times&\times&\times\\
0&\times&\times&\times
\end{array}
\right),
\ea
where the stars and crosses have the same meaning as given in 
eq.(\ref{majorana-lsnd}). However, here we have considered a set up 
in which the sterile neutrino has a mass around a few keV. We also 
make sure that the active-sterile mixing is very small, and within
the valid range given by different X-ray experiments 
\cite{Abazajian:2001vt, Boyarsky:2006fg, Boyarsky:2006zi, Loewenstein:2008yi, 
RiemerSorensen:2006pi, Hundi:2011et}. A very rough 
bound on the active-sterile mixing angle can be written as
\cite{Shaposhnikov}
\ba
\label{mass-mixing}
\theta_{14}^{2}\leq 1.8\times 10^{-5}
\left(\frac{1 ~{\rm keV}}{M_{N}^{R}}\right)^{5},
\label{x-ray-constraint}
\ea
where $M_{N}^{R}$ represents the Majorana mass of the right handed sterile 
neutrino. Therefore, we can treat the right handed neutrino as a decoupled 
state and work with the effective $3\times 3$ matrix of the active Majorana 
neutrinos. Note that the (11) element of this $3\times 3$ neutrino mass matrix 
in the basis $(\nu_{e}^{\prime} ,\nu_{\mu},\nu_{\tau})$ receives tree level 
as well as one-loop level contributions whereas the other entries in this mass 
matrix comes only through various loop corrections. The size of this tree level contribution to $(m_\nu)_{11}$ is controlled by the model parameters and for 
suitable choices of the parameters one can obtain a tree level value 
$(m_{\nu})_{\textrm {Tree}} \lsim$ 0.1 eV. Combining with the loop 
contributions one can then perform a fit to the three flavor global neutrino 
data. 

However, if we wish the keV sterile neutrino to be a candidate for 
dark matter then it should have the correct relic density 
($\Omega_N h^2 \sim$ 0.1) and must satisfy the constraints coming from 
X-ray experiments. An approximate formula for the relic density of sterile 
neutrinos via the Dodelson-Widrow (DW)\cite{Dodelson:1993je} 
mechanism is \cite{Dodelson:1993je,Abazajian:2001vt}
\ba
\Omega_N h^2 \approx 0.3 \left(\dfrac{\sin^2 2\theta}{10^{-10}} \right) 
\left(\dfrac{M^R_N}{100 ~{\rm keV}} \right)^2,
\label{sterile-nu-relic-density}
\ea
where $\Omega_N$ is the ratio of density of sterile neutrinos to the total 
density of the universe and the present value of $h$ is 0.673 \cite{planck}. 
Our numerical scan of the parameter space shows that the correct relic density 
can be achieved only if $(m_\nu)_{\rm Tree}$ is extremely small ($\sim 10^{-4}$ eV or so). 

Sterile singlet neutrino dark matter can also be produced via the resonant 
production mechanism \cite{Shi:1998km}. Several other model dependent 
production mechanisms have been discussed in the literature 
\cite{Kusenko:2006rh, Petraki:2007gq, Falkowski:2011xh, Merle:2013wta, 
Merle:2013gea}. However, in this work we assume that the relic abundance of 
keV sterile neutrinos is determined solely by 
eq.(\ref{sterile-nu-relic-density}) resulting from DW mechanism.

There have been different experimental observations which put lower 
limits on the mass of the keV warm dark matter. For fermionic dark matter
particles, a very robust lower bound on their mass comes from Pauli exclusion
principle. By demanding that the maximal (Fermi) velocity of the degenerate
fermionic gas in the dwarf spheroidal galaxies is less than the escape velocity
leads to a lower bound on the mass of the sterile neutrino dark matter 
$M_N^R >$ 0.41 keV \cite{Boyarsky:2008ju}.
This is the only model independent mass bound which holds for any fermionic 
dark matter.

Model dependent bounds such as the ones coming from phase space 
density considerations have put strong lower bounds on the mass of the sterile 
neutrino acting as a warm dark matter candidate
\cite{Boyarsky:2008ju, Rubakov, Boyarsky:2012rt}.
The authors of \cite{Seljak:2006qw, Boyarsky} put a more stringent 
lower bound on the warm dark matter mass ($M_N^R >$ 8--14 keV) by analyzing 
Lyman-$\alpha$ experimental data. In the context of left-right symmetric 
model a lower bound of $1.6$ keV on the mass of the sterile neutrino warm 
dark matter has been discussed in \cite{Lindner}. 
In Ref. \cite{Nemevsek:2012cd}, a lower limit of 0.5 keV on the sterile 
neutrino dark matter mass has been advocated in low scale left-right theory. 
In the present work we shall stick to the model independent lower bound of 
0.4 keV as discussed above. Moreover, our parameter choices are such that 
the active sterile neutrino mixing is within the valid range of experimental 
observations.

In order to get some idea about the numbers involved let us take two
examples. With a choice of $M_{1}^{D} = 805$ GeV, $M_{2}^{D} = 800$ GeV,
the R-symmetry breaking order parameter $m_{3/2} = 5$ GeV,
$\tan\beta = 5.5$, $\lambda_{S}=10^{-4}$, $f=1.5\times 10^{-4}$ one produces
a tree level mass of the active Majorana neutrino $(m_{\nu})_{\textrm {Tree}} 
\simeq 2.06\times 10^{-4}$  eV, and a sterile neutrino of mass around 0.47 keV. The active-sterile mixing is close to $4.35\times 10^{-7}$, which is within the
acceptable limit as observed by different X-ray experiments and the relic
density of the sterile neutrinos comes out to be $\Omega_N h^2 = 0.117$.
Again with another set of parameters such as $M_{1}^{D} = 1200.001$ GeV,
$M_{2}^{D} = 1200$ GeV, $m_{3/2} = 5$ GeV, $\tan\beta = 5$,
$\lambda_{S}=1.2$ and $f=1.55\times 10^{-4}$, we obtain a sterile neutrino
of $0.42$ keV mass and the tree level active Majorana neutrino mass
$(m_{\nu})_{\textrm {Tree}} \simeq 2.2\times 10^{-4}$ eV with an active-sterile mixing 
$3.64\times 10^{-7}$ and $\Omega_N h^2 = 0.114$.

To conclude this section, we observe that the keV sterile neutrino in this model
fits the requirements of a good candidate for warm dark matter.
With this note we shall now discuss different loop contributions to
the neutrino mass matrix, which provide Majorana masses for the light active 
neutrinos with appropriate mixing between them. 
\section{One loop effects to generate neutrino mass}
\label{loops}
In our model only the electron neutrino acquires a mass at the 
tree level. The other two neutrinos obtain their masses via one loop diagrams.
At one loop level, the neutrino masses are generated from diagrams involving 
charged lepton-slepton loop, quark-squark loop and neutralino-Higgs loop 
respectively. We note in passing that similar onle loop calculations 
have also been performed in \cite{Claudia}, which fits neutrino masses via 
radiative corrections only, without introducing an extra right handed 
neutrino superfield. 
\subsection{Charged lepton-slepton loop}
\label{lepton-slepton}
We first consider the charged lepton-slepton
loop which will generate Majorana mass terms for the 
neutrinos of all flavors \cite{Yuval}. We consider only the tau-stau
loop as other charged lepton-slepton loops have very 
mild effect as far as neutrino mass is concerned.
\begin{figure}[htb]
\begin{center}
\includegraphics[height=1.8in,width=2.8in]{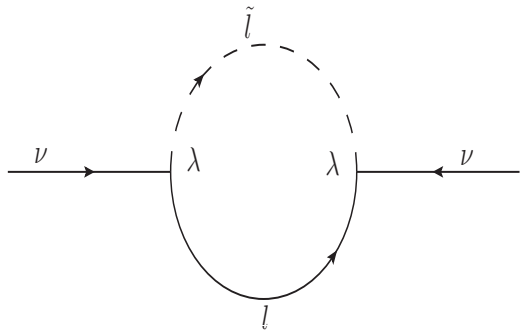} 
\caption{Charged lepton-slepton loop}
\end{center}
\label{stau-tau-loop}
\end{figure}
The contribution of the stau-tau loop (see, fig.\ref{stau-tau-loop}) to the one loop neutrino mass matrix is 
\ba
m_{\nu}^{l-s}&=&\frac{1}{(16\pi^{2})^{2}}
\left[\frac{m_{\tau}m_{3/2}v_{a}}{m_{\tilde \tau}^{2}}\right]
\hat\beta_{\tau}\left(
\begin{array}{ccc}
\lambda_{133}^{2} & \lambda_{133}\lambda_{233} & 0\\
\lambda_{233}\lambda_{133} & \lambda_{233}^{2} & 0\\
0 & 0 & 0
\end{array}
\right)\ln\left(\frac{m_{{\tilde \tau}_1}^{2}}{m_{{\tilde \tau}_2}^{2}}\right),
\ea
where we have used the expression of $A_\tau$ from eq.(\ref{amsb-a-terms}), 
which provide the necessary lepton number violation of two units in the scalar 
propagator. Here $m_{{\tilde \tau}_2}^{2} > m_{{\tilde \tau}_1}^{2}$ represent 
the physical squared-masses of the staus and $m_{{\tilde \tau}}^{2} 
\simeq m_{{\tilde \tau}_2}^{2}$. In the above mass matrix we considered 
$e=1$, and $\mu,\tau=2,3$ respectively, keeping in mind that $\lambda$ is 
antisymmetric in the first two indices. Because of this antisymmetry property 
of the coupling $\lambda$, some of the elements in $m_{\nu}^{l-s}$ 
are zero.
\subsection{Squark-quark loop}
\label{squark-quark}
\begin{figure}[htb]
\centering
\includegraphics[height=1.8in, width=2.8in]{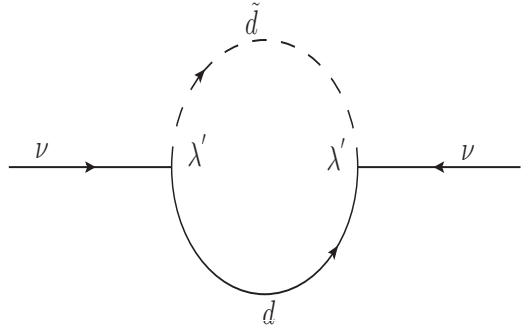}
\caption{Quark-squark loop for $d =b,s$ quarks and 
${\tilde d} = {\tilde b}, ~{\tilde s}$ squarks.}
\label{squark-quark-loop}
\end{figure}

The squark-quark loop will also contribute to the light neutrino 
Majorana mass matrix \cite{Yuval}. Here we have taken into account bottom and 
strange squark-quark loop as shown in fig.(\ref{squark-quark-loop}). 
The contribution of quark-squark loop to the one loop neutrino mass matrix is
\ba
\label{quark-squark}
&&m_{\nu}^{q-s}= \frac{3}{(16\pi^{2})^{2}}\left
[\frac{m_{3/2}v_{a}}{m_{\tilde b}^{2}}\right]
\hat\beta_{b}\nonumber \\
&&\times \left[m_{b}\left(
\begin{array}{ccc}
\lambda_{133}^{\prime 2} & \lambda_{133}^{\prime}\lambda_{233}^{\prime}&
\lambda_{133}^{\prime}\lambda_{333}^{\prime}\\
\lambda_{133}^{\prime}\lambda_{233}^{\prime} & \lambda_{233}^{\prime 2}&
\lambda_{233}^{\prime}\lambda_{333}^{\prime}\\
\lambda_{333}^{\prime}\lambda_{133}^{\prime}& \lambda_{333}^{\prime}
\lambda_{233}^{\prime} & \lambda_{333}^{\prime 2}
\end{array}
\right)
\ln\left(\frac{m_{{\tilde b}_1}^{2}}{m_{{\tilde b}_2}^{2}}\right)\right.
+\left. m_{s}\left(
\begin{array}{ccc}
0 & 0 & 0\\
0 & \lambda_{223}^{\prime}\lambda_{232}^{\prime} & \lambda_{223}^{\prime}
\lambda_{332}^{\prime}\\
0 & \lambda_{323}^{\prime}\lambda_{232}^{\prime} & \lambda_{323}^{\prime}
\lambda_{332}^{\prime}
\end{array}
\right)\ln\left(\frac{m_{{\tilde b}_1}^{2}}{m_{{\tilde b}_2}^{2}}\right)
\right]\nonumber \\
&+&\frac{3}{(16\pi^{2})^{2}}\left[\frac{m_{3/2}v_{a}}{m_{\tilde s}^{2}}
\right]\hat\beta_{s}\nonumber \\
&&\times \left[m_{b}\left(
\begin{array}{ccc}
0&0&0\\
0&\lambda_{232}^{\prime}\lambda_{223}^{\prime}
&\lambda_{232}^{\prime}\lambda_{323}^{\prime}\\
0&\lambda_{332}^{\prime}\lambda_{223}^{\prime}
&\lambda_{332}^{\prime}\lambda_{323}^{\prime}
\end{array}
\right)\ln\left[\frac{m_{{\tilde s}_1}^{2}}{m_{{\tilde s}_2}^{2}}\right]
+m_{s}\left(
\begin{array}{ccc}
\lambda_{122}^{\prime 2}&\lambda_{122}^{\prime}\lambda_{222}^{\prime}
&\lambda_{122}^{\prime}\lambda_{322}^{\prime}\\
\lambda_{122}^{\prime}\lambda_{222}^{\prime}&\lambda_{222}^{\prime2}
&\lambda_{222}^{\prime}\lambda_{322}^{\prime}\\
\lambda_{322}^{\prime}\lambda_{122}^{\prime}&\lambda_{322}^{\prime}
\lambda_{222}^{\prime}&\lambda_{322}^{\prime 2}
\end{array}
\right)\ln\left[\frac{m_{{\tilde s}_1}^{2}}{m_{{\tilde s}_2}^{2}}\right]
\right].  \nonumber \\
\ea
$m_{{{\tilde b}_1},{{\tilde b}_2}}^2$ $m_{{{\tilde s}_1},{{\tilde s}_2}}^2$ 
are the physical squared-masses of the sbottom squarks and strange squarks 
respectively with $m_{{\tilde b}_2}^2 > m_{{\tilde b}_1}^2$, 
$m_{{\tilde s}_2}^2 > m_{{\tilde s}_1}^2$ and $m_{{\tilde b}}^2 
\simeq m_{{\tilde b}_2}^2$, $m_{{\tilde s}}^2 \simeq m_{{\tilde s}_2}^2$.
Since $\hat\beta_{b}\gg\hat\beta_{s}$ and $m_{b}\gg m_{s}$,
the dominant contribution to the neutrino mass matrix arises from the first 
two terms in eq.(\ref{quark-squark}) as long as we assume 
$m^2_{\tilde s} \gg m^2_{\tilde b}$. The other terms have a sub dominant 
contribution to the neutrino masses and therefore, it is safe 
to consider only the first two terms for computing the neutrino mass 
eigenvalues. 
\subsection{Neutralino-Higgs boson loop}
\label{Neutrino-neutralino-sneutrino}
\begin{figure}[htb]
\centering
\includegraphics[height=1.8in, width=2.8in]{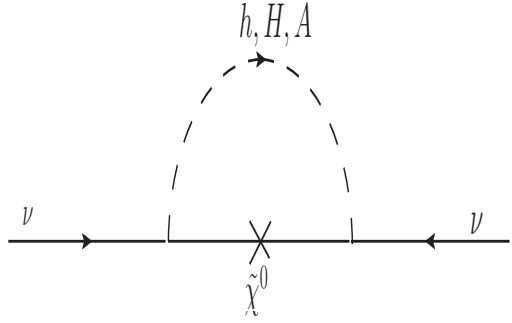}
\caption{Neutralino-Higgs boson loop}
\label{neutralino-higgs-loop}
\end{figure}

We now consider the loop consisting of neutralino and 
Higgs propagators to generate Majorana mass of neutrinos 
\cite{Davidson, Davidson-1}.
This is shown in fig.\ref{neutralino-higgs-loop}. 
The loop contribution is proportional to the Majorana gaugino
mass parameters $M_1$ and $M_2$, which are much smaller than 
the mass of the corresponding physical neutralino states.
Note that the bilinear term $b\mu_L H_u^{0}{\tilde \nu}^a$
in the scalar potential can be large in this model and this loop
contribution can be significant. In order to compute this loop we 
consider a simplified scenario where the singlet and the triplet states 
are integrated out and we are therefore left only with the $H_u$ and 
${\tilde \nu}^a$ fields as considered earlier in
the discussion of the scalar sector, generating the CP-even physical states
$h, H$ and the CP-odd state $A$. 

Majorana mass term of a neutrino implies lepton number violation 
by two units. This is provided by the Majorana mass insertion in the 
neutralino propagator. The contribution to the neutrino mass matrix from 
this loop is given by
\ba
&&(m_{\nu})_{11}=
\frac{g^{2}}{64\pi^{2}}\sum_{\gamma=1,2}
\left[Z_{\gamma 2}-\tan\theta_{W}Z_{\gamma 1}\right]^{2}
\frac{M_{1}}{2}\nonumber \\
&&\left[\cos^2\alpha B_{0}(0,m_{H}^{2},m_{\tilde\chi^{0}}^{2})
+\sin^2\alpha B_{0}(0,m_{h}^{2},m_{\tilde\chi^{0}}^{2})-
\sin^2\beta B_{0}(0,m_{A}^{2},m_{\tilde\chi^{0}}^{2})\right]\nonumber \\
&&+\frac{g^{2}}{64\pi^{2}}\sum_{\gamma=3,4}
\left[Z_{\gamma 2}-\tan\theta_{W}Z_{\gamma 1}\right]^{2}
\frac{M_{2}}{2}\nonumber \\
&&\left[\cos^2\alpha B_{0}(0,m_{H}^{2},m_{\tilde\chi^{0}}^{2})
+\sin^2\alpha B_{0}(0,m_{h}^{2},m_{\tilde\chi^{0}}^{2})-
\sin^2\beta B_{0}(0,m_{A}^{2},m_{\tilde\chi^{0}}^{2})\right],
\label{eq-neutralino-higgs}
\ea
where $\tan\beta = v_u/v_a$ and we have used
\ba
{\tilde \nu}_R^a &\simeq& v_1 + 
\frac{1}{\sqrt 2}(H \cos\alpha 
- h \sin\alpha), ~~~~~(a = 1(e)) \nonumber \\
H^0_u &\simeq& v_2 +
\frac{1}{\sqrt 2} (H \sin\alpha + h \cos\alpha), \nonumber \\
{\tilde \nu}_I^a &\simeq& \frac{1}{\sqrt 2}(G \cos\beta + A \sin\beta).
\ea
The summation in eq.(\ref{eq-neutralino-higgs}) is taken over 
two pairs of nearly degenerate pseudo-Dirac heavier neutralino states 
$m_{{\tilde \chi}_{1,2}}$ and $m_{{\tilde \chi}_{3,4}}$, which are 
predominantly bino (${\tilde b}^0$) and wino (${\tilde w}^0$) respectively.
Here we have assumed that $|m_{{\tilde \chi}_{1,2}}| \simeq M^D_1 \pm 
\dfrac{M_1}{2}$ and $|m_{{\tilde \chi}_{3,4}}| \simeq M^D_2 \pm \dfrac{M_2}{2}$ and for a given pair the neutralino mixing matrix elements $Z_{\gamma 2}$ 
and $Z_{\gamma 1}$ does not change for $\gamma = (1,2) ~{\rm and}~ (3,4)$. 
$B_{0}$ is a Passarino-Veltman function and follow its definition as mentioned 
in \cite{Yuval,Davidson,Davidson-1}. It is important to note that this one loop contribution adds only to the (11) element of the effective 3$\times$3 neutrino mass matrix. The other neutrino flavors do not get any contribution
to their masses from this loop because the corresponding sneutrinos do not 
mix with $H_u$.
\section{Numerical analysis}
\label{numerical-analysis}
We now present the results of our detailed numerical investigations 
to fit the lightest Higgs boson mass, neutrino masses and mixing 
angles as well as the keV sterile neutrino mass and its mixing with 
the active neutrino. As mentioned earlier in the text, we analyze two 
situations, one with small singlet and triplet couplings ($\lambda_S$ 
and $\lambda_T$ respectively), which would imply heavy stops to fit 
the lightest Higgs boson mass whereas the other case with light stop 
mass requires large $\lambda_S$ and $\lambda_T$, which would provide 
significant radiative corrections to the lightest Higgs boson mass. 
A set of benchmark points for the latter case is provided below in 
Table \ref{Table-2}.

\begin{table}[ht]
\centering
\begin{tabular}{|c| c| c| c|}
\hline \hline
Parameters & BP-1 & BP-2 & BP-3\\
\hline
$M_{1}^{D}$ & 1200.001 GeV& 1000.001 GeV & 800.001 GeV\\
$M_{2}^{D}$ & 1200 GeV& 1000 GeV& 800 GeV\\
$\tan\beta$ & 5& 7 & 10\\
$\lambda_{S}$& 1.25& 1.1 & 0.98\\
$\lambda_{T}$& $\lambda_{S}\tan\theta_{W}\sim$ 0.69&
0.6 & 0.54\\
$\mu$ & 590 GeV& 530 GeV & 650 GeV\\
$t_S$ & $(200)^{3}$ (GeV)$^{3}$& $(200)^{3}$ (GeV)$^{3}$ 
& $(200)^{3}$ (GeV)$^{3}$\\
b$\mu_{L}$ & $-(200)^{2}$ (GeV)$^{2}$& $-(200)^{2}$ (GeV)$^{2}$
 & $-(200)^{2}$ (GeV)$^{2}$\\
$m_{S}$ & 7.6 TeV & 10 TeV & 18 TeV\\
$m_{T}$ & 5.46 TeV & 5.8 TeV& 1.9 TeV\\
$v_{S}$ & -0.6 GeV & -0.3 GeV & -0.1 GeV\\
$v_{T}$ & 0.1 GeV & 0.1 GeV & 0.05 GeV\\
f & $1.55\times10^{-4}$ & $1.1\times10^{-4}$ & $1.0\times10^{-4}$\\
$M_{R}$ &  3.67 GeV& 3 GeV & 3.16 GeV\\
$m_{\tilde t_{1}}=m_{\tilde t_{2}}$ & 600 GeV & 900 GeV & 1.2 TeV\\
$b_{S}$ & 1 TeV & 1 TeV & 1 TeV\\
$b_{T}$ & 1 TeV & 1 TeV & 1 TeV\\
$m_{3/2}$ & 5 GeV & 6 GeV & 3 GeV\\
\hline
$m_{h}$ & 125.15 GeV & 124.9 GeV & 123.7 GeV\\
$M_{N}^{R}$ & 0.42 keV & 0.51 keV & 0.43 keV\\
$(m_{\nu})_{\textrm{Tree}}$ & $2.17\times 10^{-4}$ eV & $1.86\times 10^{-4}$ eV & $2.4\times 10^{-4}$ eV\\
$\theta_{14}^{2}$ & 5.05$\times 10^{-7}$ & 3.64$\times 10^{-7}$ 
& $5.53\times10^{-7}$ \\ 
$\Omega_{N}h^{2}$ & 0.1121 & 0.114 & 0.122 \\
\hline
\end{tabular}
\caption{Benchmark points (with large $\lambda_S$ and $\lambda_T$) to calculate the lightest Higgs boson mass, light active neutrino mass, mass of the sterile 
neutrino as well as its mixing with active neutrino and the relic density of 
sterile neutrino dark matter.}
\label{Table-2}
\end{table}

We chose a relatively larger value of the Dirac wino mass 
consistent with the allowed range of $\tan\beta$ ($2.7\leq \tan\beta\leq 17.4$) 
obtained from the deviation in the couplings of the Z boson to charged leptons 
as well as from the $\tau$ Yukawa couplings \cite{Ponton}. In order to fit the 
neutrino data, we choose the Dirac bino mass very close to the Dirac wino mass. Let us emphasize that in this model a large Dirac gaugino mass does not 
introduce any logarithmically divergent contribution to scalar masses squared 
because it is cancelled by the new scalar loop contributions \cite{Fox-Weiner}.  

In Table \ref{Table-3} we show benchmark points corresponding to  
small $\lambda_{T}\sim 10^{-4}$. In this case, in order to fit the neutrino 
data, one does not require a strong degeneracy between $M_1^D$ and $M^D_2$. 
It is worth mentioning once again that we have reduced the number of 
independent parameter of the model by assuming certain relations between 
some of them as shown in eqs.(\ref{relation-MR-f}) 
and (\ref{relation-lamt-lams}).
\begin{table}[ht]
\centering
\begin{tabular}{|c| c| c| c|}
\hline \hline
Parameters & BP-4 & BP-5 & BP-6\\
\hline
$M_{1}^{D}$ & 1018 GeV& 805 GeV & 604 GeV\\
$M_{2}^{D}$ & 1000 GeV& 800 GeV& 600 GeV\\
$\tan\beta$ & 10& 5.5 & 7\\
$\lambda_{S}$& $10^{-4}$& $10^{-4}$ & $10^{-4}$\\
$\lambda_{T}$& $\lambda_{S}\tan\theta_{W}\sim$ $5.5\times 10^{-5}$&
$5.5\times 10^{-5}$ & $5.5\times 10^{-5}$\\
$\mu$ & 700 GeV& 500 GeV & 580 GeV\\
$t_S$ & $(200)^{3}$ (GeV)$^{3}$& $(200)^{3}$ (GeV)$^{3}$ 
& $(200)^{3}$ (GeV)$^{3}$\\
b$\mu_{L}$ & $-(200)^{2}$ (GeV)$^{2}$& $-(200)^{2}$ (GeV)$^{2}$
 & $-(200)^{2}$ (GeV)$^{2}$\\
$m_{S}$ & 12 TeV & 11.6 TeV & 11 TeV\\
$m_{T}$ & 11 TeV & 10.14 TeV& 9 TeV\\
$v_{S}$ & -0.1 GeV & -0.1 GeV & -0.1 GeV\\
$v_{T}$ & 0.1 GeV & 0.1 GeV & 0.1 GeV\\
f & $0.92\times 10^{-4}$ & $1.5\times 10^{-4}$ & $1.2\times 10^{-4}$\\
$M_{R}$ &  3.69 GeV& 2.62 GeV & 2 GeV\\
$m_{\tilde t_{1}}=m_{\tilde t_{2}}$ & 6.5 TeV & 6.5 TeV & 6.5 TeV\\
$b_{S}$ & 1 TeV & 1 TeV & 1 TeV \\
$b_{T}$ & 1 TeV & 1 TeV & 1 TeV \\
$m_{3/2}$ & 3.5 GeV & 5 GeV & 6 GeV\\
\hline
$m_{h}$ & 126 GeV & 123.1 GeV & 124.9 GeV\\
$M_{N}^{R}$ & 0.41 keV & 0.47 keV & 0.59 keV\\
$(m_{\nu})_{\textrm{Tree}}$ & $2.41\times 10^{-4}$ eV & $2.06\times 10^{-4}$eV & $1.6\times 10^{-4}$ eV\\
$\theta_{14}^{2}$ & 5.85$\times 10^{-7}$ & 4.35$\times 10^{-7}$ 
& $2.7\times10^{-7}$ \\ 
$\Omega_{N}h^{2}$ & 0.119 & 0.117 & 0.114\\
\hline
\end{tabular}
\caption{Benchmark points with small $\lambda_S$ and $\lambda_T$.} 
\label{Table-3}
\end{table}
One can observe from these two tables that the benchmark points 
provide a lightest Higgs boson mass around 125 GeV, a sterile neutrino mass in 
the keV range along with a very small active-sterile mixing and 
a very small tree level active neutrino Majorana mass. The mass and mixing 
of the sterile neutrino are in the allowed range of values coming from X-ray 
observations and it can be accommodated as a warm dark matter candidate in our 
model. 
\begin{figure}[ht!]
\begin{center}
\includegraphics[width=0.4\textwidth]{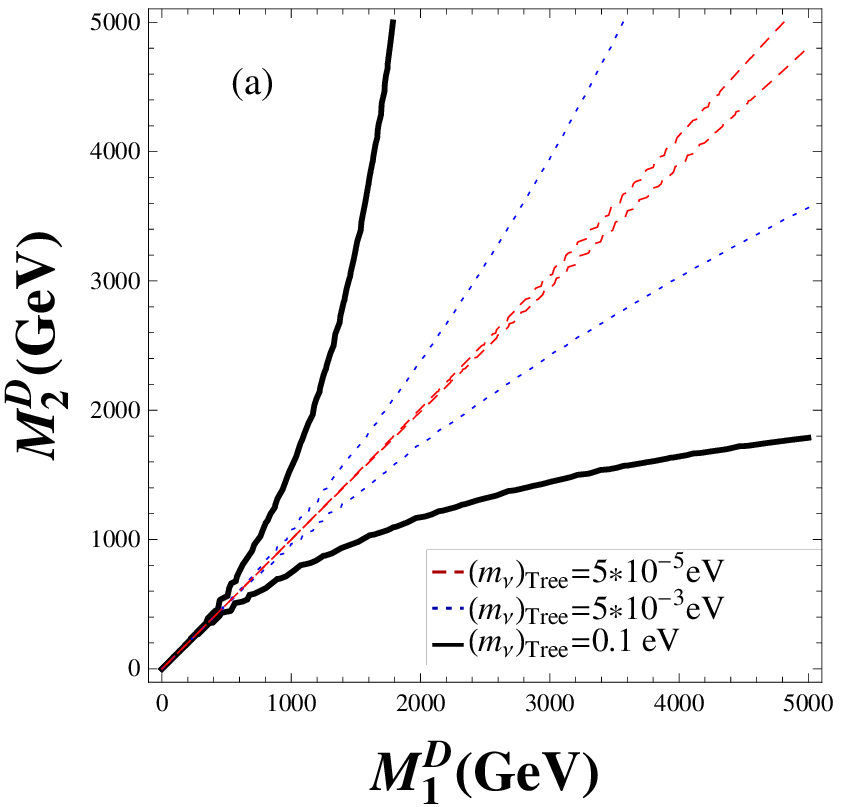}
{%
\includegraphics[width=0.4\textwidth]{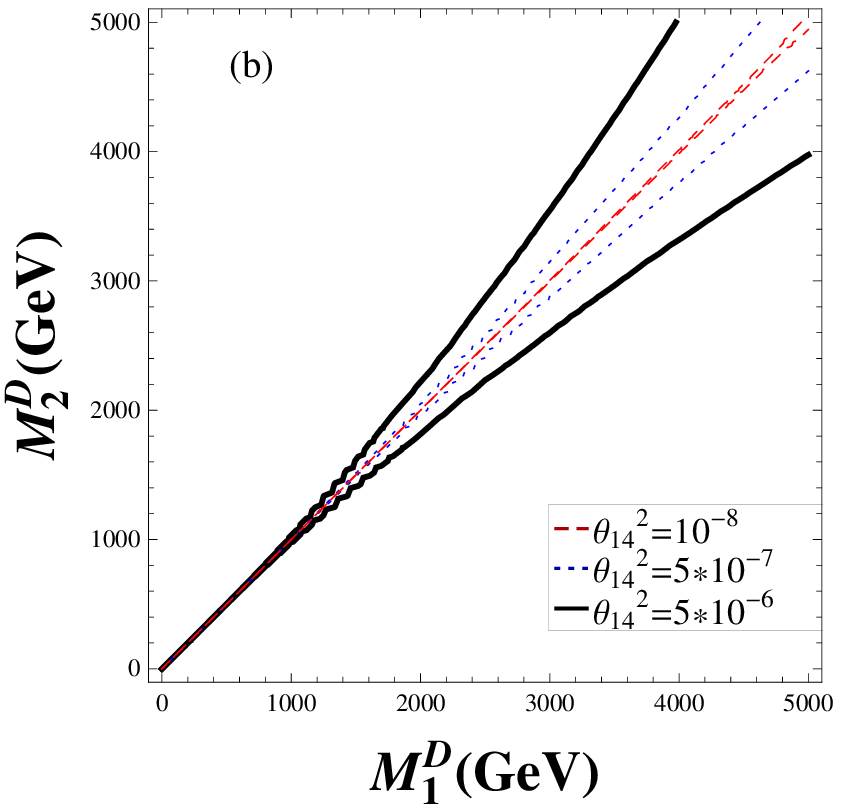}
}\\ 
{%
\includegraphics[width=0.4\textwidth]{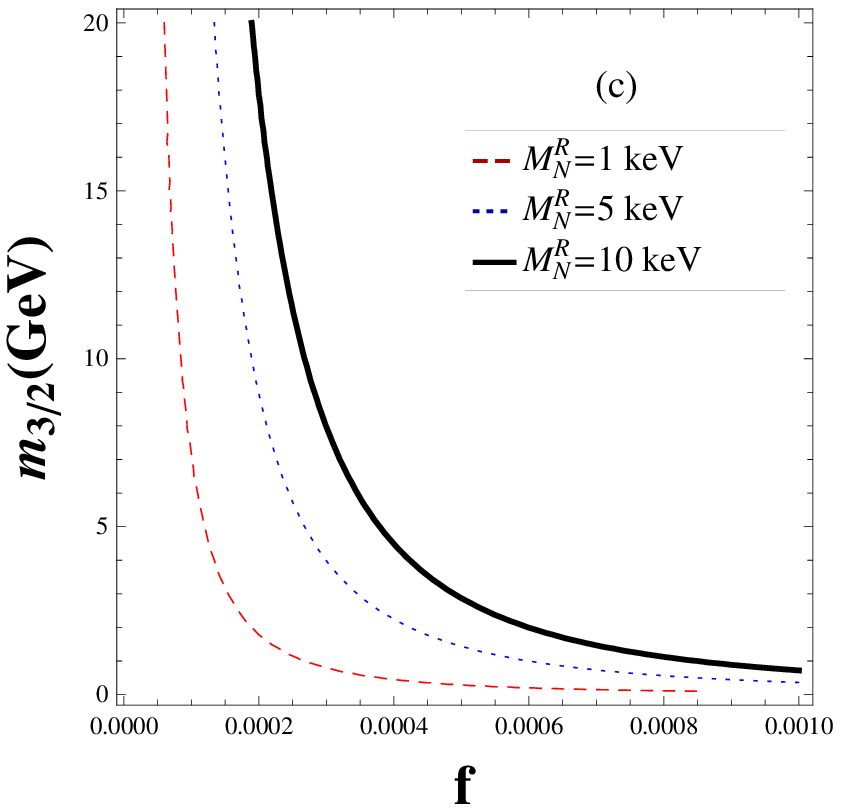}
}%
{%
\includegraphics[width=0.4\textwidth]{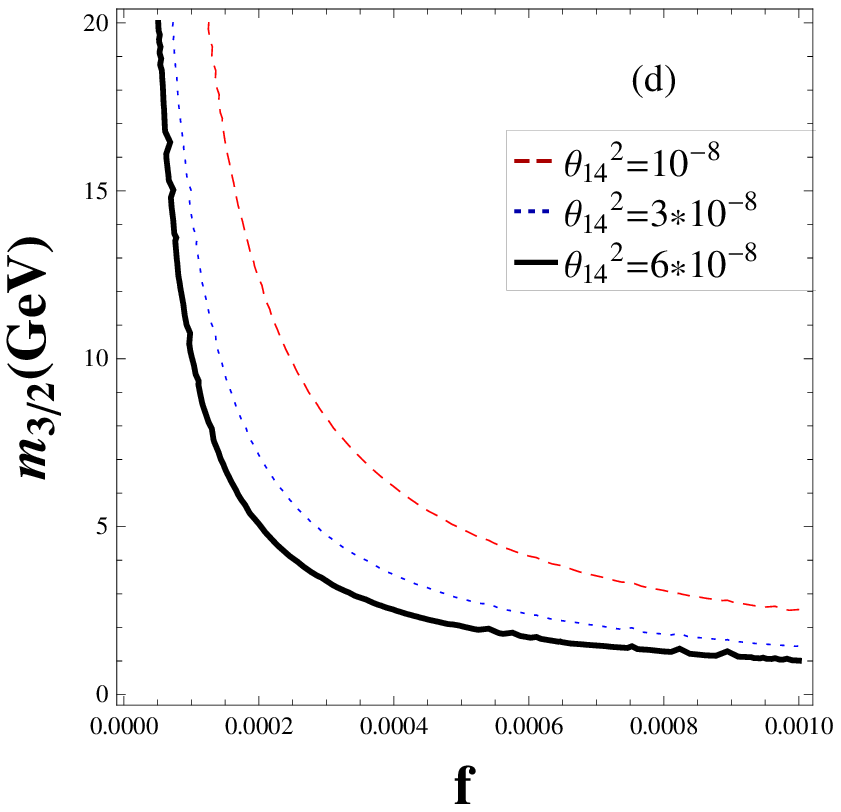}
}%
\end{center}
\caption{
Figure (a) represents contours of tree level light neutrino mass 
$(m_{\nu})_{\textrm {Tree}}$ in the ($M_{1}^{D}$--$M_{2}^{D}$) plane. The black thick line represents $(m_\nu)_{\textrm {Tree}}$ = 0.1 eV. The blue dotted line and the red dashed line represent $(m_\nu)_{\textrm {Tree}}$ = $5\times 10^{-3}$ eV and $5\times 10^{-5}$ eV, respectively. Figure (b) represents 
active-sterile mixing ($\theta^2_{14}$) in the same ($M_{1}^{D}$--$M_{2}^{D}$) 
plane. The black thick line represents the contour of $5\times 10^{-6}$ and 
the blue dotted line and the red dashed line represent contours of 
$5\times 10^{-7}$ and $10^{-8}$, respectively. Figures (c) and (d) show 
the contours of sterile neutrino mass ($M^R_N$) and $\theta^2_{14}$ in the 
($f$--$m_{3/2}$) plane. In figure (c) the black thick line corresponds to 
$M^R_N$ = 10 keV whereas the blue dotted line and the red dashed line show 
contours of $M^R_N$ = 5 keV and 1 keV, respectively. In figure (d) the black 
thick line shows a mixing of $6\times 10^{-8}$ and the blue and the red line 
show $\theta^2_{14}$ = $3\times 10^{-8}$ and $10^{-8}$, respectively.}
\label{fig1:subfigures}
\end{figure}

\begin{figure}[ht!]
\begin{center}
%
{%
\includegraphics[width=0.4\textwidth]{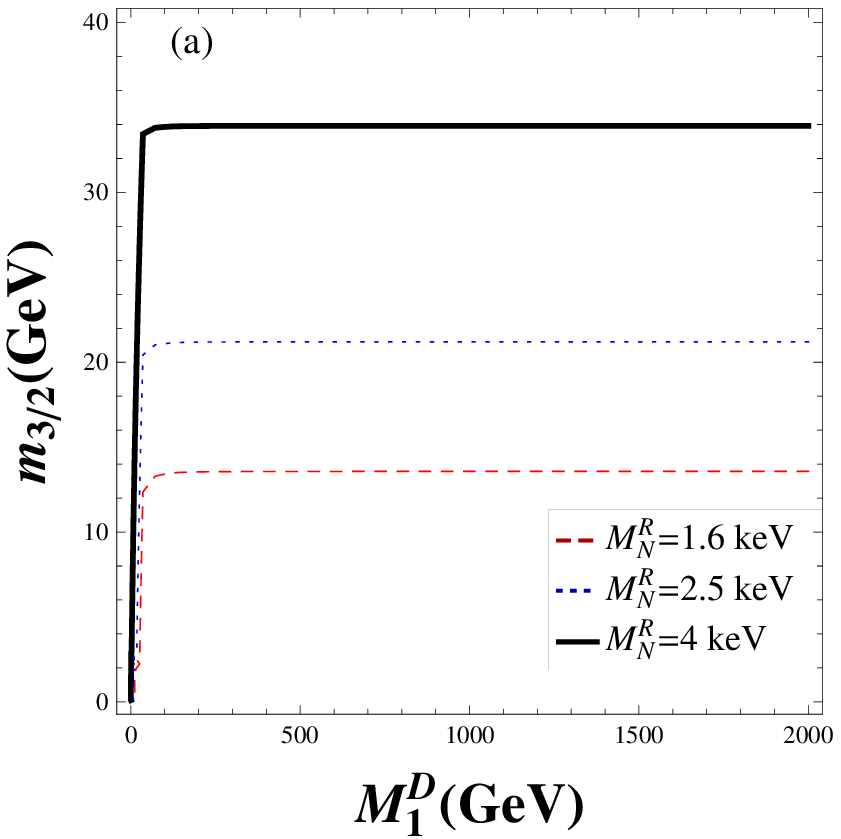}
}%
{%
\includegraphics[width=0.4\textwidth]{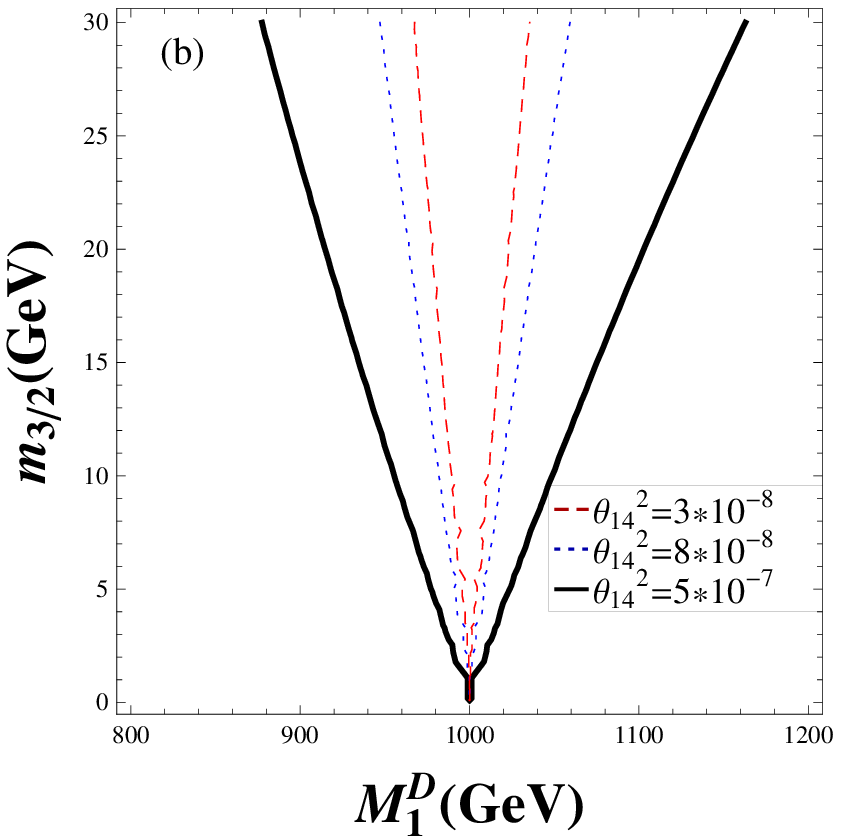}
}\\ 
{%
\includegraphics[width=0.4\textwidth]{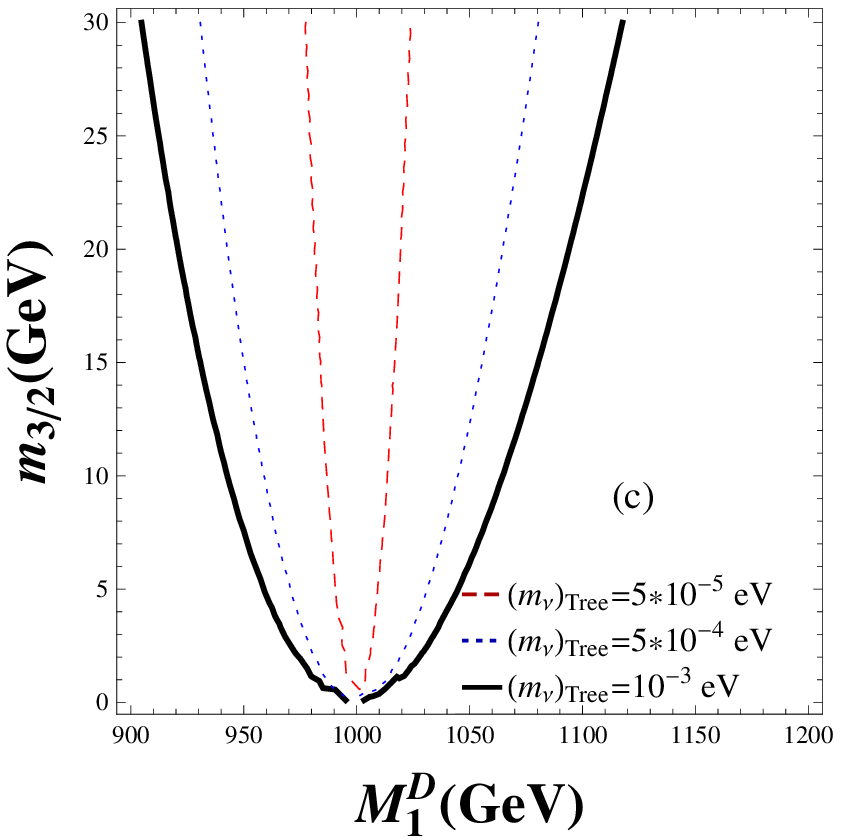}
}%
{%
\includegraphics[width=0.4\textwidth]{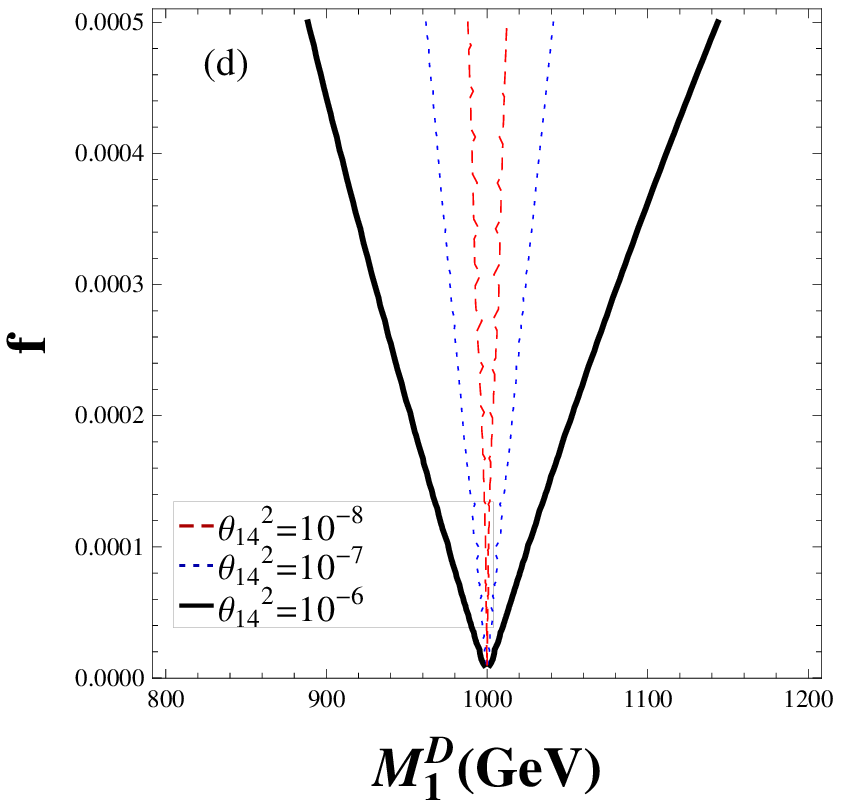}
}%
\end{center}
\caption{
Figure (a) represents contours of $M_N^R$ in the ($M_{1}^{D}$--$m_{3/2}$) 
plane. The red dashed line shows a sterile neutrino of mass 1.6 keV whereas 
the blue dotted and the thick black line shows a sterile neutrino mass of 
2.5 keV and 4 keV, respectively. In figure (b) $\theta^2_{14}$ contours are 
shown in the same plane. The contours are $5\times 10^{-7}$ (black-thick), 
$8\times 10^{-8}$ (blue-dotted) and $3\times 10^{-8}$ (red-dashed), 
respectively. In figure (c) we show the variation of tree level active neutrino mass $m_\nu$ in the ($M_{1}^{D}$--$m_{3/2}$). The outermost contours represent 
$(m_\nu)_{\textrm {Tree}}$ = $10^{-3}$ eV. Finally in figure (d) we plot the 
contours of $\theta^2_{14}$ in the ($M_{1}^{D}$--$f$) plane.  
}%
\label{fig2:subfigures}
\end{figure}

In figure \ref{fig1:subfigures}(a) the contours of the tree level mass 
$(m_{\nu})_{\textrm {Tree}}$ of the light active neutrino in the 
($M_{1}^{D}$-$M_{2}^{D}$) plane exhibits the degeneracy 
required for these two parameters in order to have a small neutrino mass. 
Figure \ref{fig1:subfigures}(b) shows that the active-sterile mixing is 
also dependent on the degeneracy of $M_{1}^{D}$ and $M_{2}^{D}$. Since the 
X-ray experiments provide very stringent constraints on the mixing, one 
is compelled to choose the Dirac gaugino masses close to each other. 
For these two plots, all the other parameters are fixed at the 
values of BP-4.
In figure \ref{fig1:subfigures}(c) we show the variation of the sterile 
neutrino mass in the ($f$-$m_{3/2}$) plane. The figure shows that for a 
fixed $f$, a larger gravitino mass produces a larger mass of the sterile 
neutrino. Again we expect this to happen because the gravitino is the order 
parameter of R-breaking and therefore, a larger gravitino mass creates a 
larger mass splitting between the sterile and the active neutrino, 
which would be zero in the absence of gravitino mass. This way the sterile 
neutrino mass gets more enhanced whereas the active neutrino mass becomes 
smaller. On the contrary the active-sterile mixing decreases with $m_{3/2}$ for a fixed $f$ as shown in figure \ref{fig1:subfigures}(d). This is also expected, as a larger gravitino mass increases the mass of the sterile neutrino and thus 
reduces its mixing with the active neutrino. In figures \ref{fig1:subfigures}
(c) and \ref{fig1:subfigures}(d), we have fixed $M_{2}^{D}$ at 1 TeV and 
$M_{1}^{D}$ at 1.018 TeV, corresponding to BP-4
in Table \ref{Table-3}.
In figure \ref{fig2:subfigures} the contours of (a) $M_N^R$, (b) 
$\theta^2_{14}$ and (c) $(m_\nu)_{\textrm {Tree}}$ are shown in 
the ($M_{1}^{D}$--$m_{3/2}$) plane and in (d) contours of
$\theta^2_{14}$ are presented in the ($f$--$M^D_1$) plane for other 
parameter choices shown in BP-4. 
One can see from figure \ref{fig2:subfigures}(a) that for large values
of $M_{1}^{D}$, the sterile neutrino mass $M_{N}^{R}$ is almost 
insensitive to $M_{1}^{D}$ as expected from eq.(\ref{Sterile-mass}).
However, the mixing $\theta^2_{14}$  increases with $M^D_1$ for a fixed 
$m_{3/2}$ and this is because of the fact that the light neutrino mass 
$m_\nu$ also increases with $M^D_1$ for a fixed $M^D_2$ and $m_{3/2}$ 
(see figure \ref{fig2:subfigures}(c)) and thus leads to an increase in 
$\theta^2_{14}$. The variation of $\theta^2_{14}$ in the ($f$--$M^D_1$) 
plane can also be explained in a similar way by looking at 
eq.(\ref{majorana-mass-tree-level}).

\begin{figure}[htb]
\centering
\includegraphics[height=3.5in, width=4.5in]{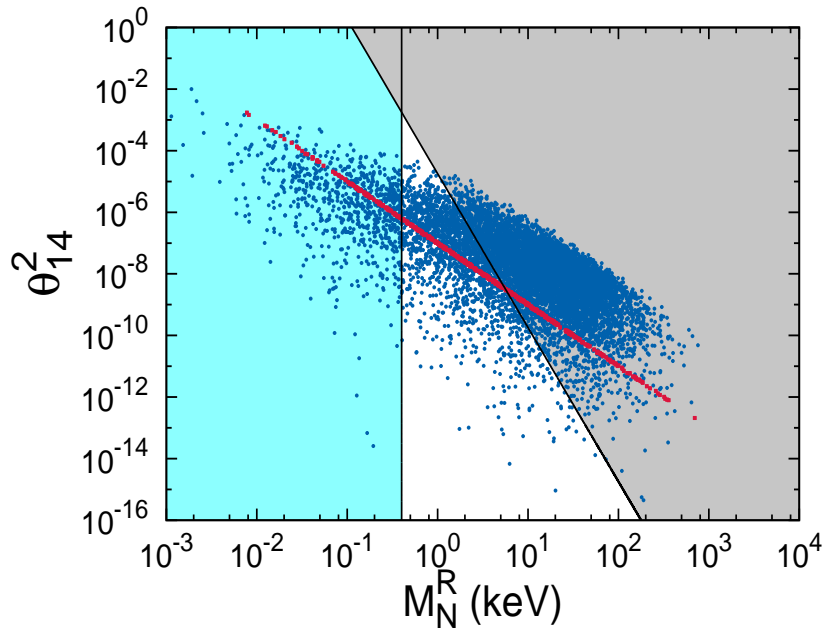}
\caption{
Scatter plot in sterile neutrino mass and active-sterile mixing plane 
showing the allowed regions, in the heavy stop scenario. The grey region shows 
the part of the parameter space excluded by the X-ray experiments. 
Lower bound on sterile neutrino mass excludes the blue region to the left
of the vertical line and the thick red band represents the parameter points 
which satisfy correct dark matter relic density, at 3$\sigma$.}
\label{sterile-mass-mixing1}
\end{figure}

\begin{figure}[t]
\centering
\includegraphics[height=3.5in, width=4.5in]{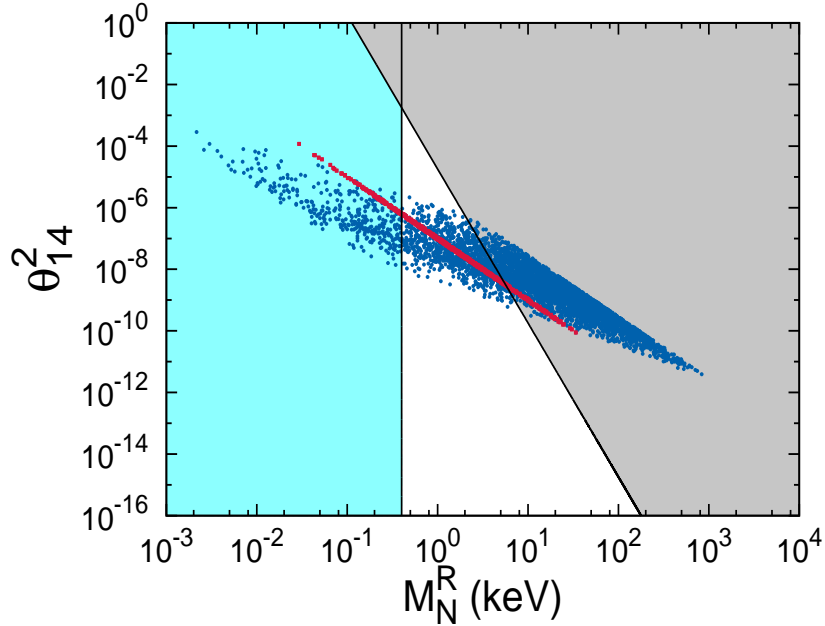}
\caption{
Scatter plot in the ($M^R_N$--$\theta^2_{14}$) plane showing the allowed 
regions, in the light stop scenario. The colored(shaded) regions are the same 
as described in figure \ref{sterile-mass-mixing1}.}
\label{sterile-mass-mixing2}
\end{figure}

We have also presented two scatter plots in the ($M_N^R$--$\theta^2_{14}$) 
plane in figures \ref{sterile-mass-mixing1} and \ref{sterile-mass-mixing2} 
showing the allowed region after taking into account the constraints from 
the X-ray experiments as well as the lower bound of 0.4 keV on the sterile 
neutrino mass, discussed earlier . On top of that we have also shown the 
points satisfying the correct dark matter relic density at 3$\sigma$ 
($\Omega_{\rm DM} h^2 = 0.1199 \pm 0.0027$ at $1\sigma$) as obtained from the 
recent observations of the PLANCK experiment \cite{planck}. 

Note that in this model the gravitino is the LSP for the parameter region 
discussed so far and can, in principle, be a candidate for dark matter. So, 
in general, one can have two component dark matter in this model. However, in 
figures \ref{sterile-mass-mixing1} and \ref{sterile-mass-mixing2} we have 
assumed that the dark matter relic density is entirely due to the sterile 
neutrino. A more detailed analysis of this two component dark matter scenario 
is beyond the scope of the present work. In order to generate 
these two plots we have varied all the parameters, which play an important role 
in sterile neutrino mass and active sterile mixing. 
This plot has been generated by varying the model parameters in the following 
range: 800 GeV $\leq M^D_1,~M^D_2 \leq$ 850 GeV, 1 GeV $\leq m_{3/2} \leq$ 40 
GeV, $10^{-5} \leq f \leq 8\times 10^{-4}$ and 2.7 $\leq \tan\beta \leq$ 17. 
We have kept $\lambda_{S} \sim 10^{-4}$ and so obviously these points represent 
the heavy stop scenario. The grey region is disallowed by the constraints
from X-ray observations whereas the red line at 0.4 keV and the 
blue region to its left is ruled out by the lower bound on sterile neutrino
mass. Finally, note that by varying the stop mass we ensured that all the 
scattered points in this plot produced the lightest Higgs boson mass in the 
range $(123-127)$ GeV.
In figure \ref{sterile-mass-mixing2}, we show the results of our parameter 
space scan in the light stop scenario. In this plot we have used $\lambda_{S} 
\sim ~1.1$ and 1 GeV $\leq  m_{3/2} \leq$ 40 GeV whereas $f$ and $\tan\beta$ 
are varied in the same range as before. 

We discussed earlier that for large $\lambda_{S}$, the Dirac 
gaugino masses $M^D_1$ and $M^D_2$ need to be almost degenerate in order to 
fit a small tree level mass of the active neutrino. Therefore in this 
plot we fixed $M_{1}^{D}$ = 1000.001 GeV and $M_{2}^{D}$ = 1000 GeV.
The grey and the blue regions again represent the parameter points ruled out 
by X-ray experiments and lower limit on the sterile neutrino mass respectively. We have also ensured that each and every point in this scattered plot produce a Higgs boson mass in the range $(123-127)$ GeV.
\begin{figure}[t]
\centering
\includegraphics[height=3.5in, width=4.5in]{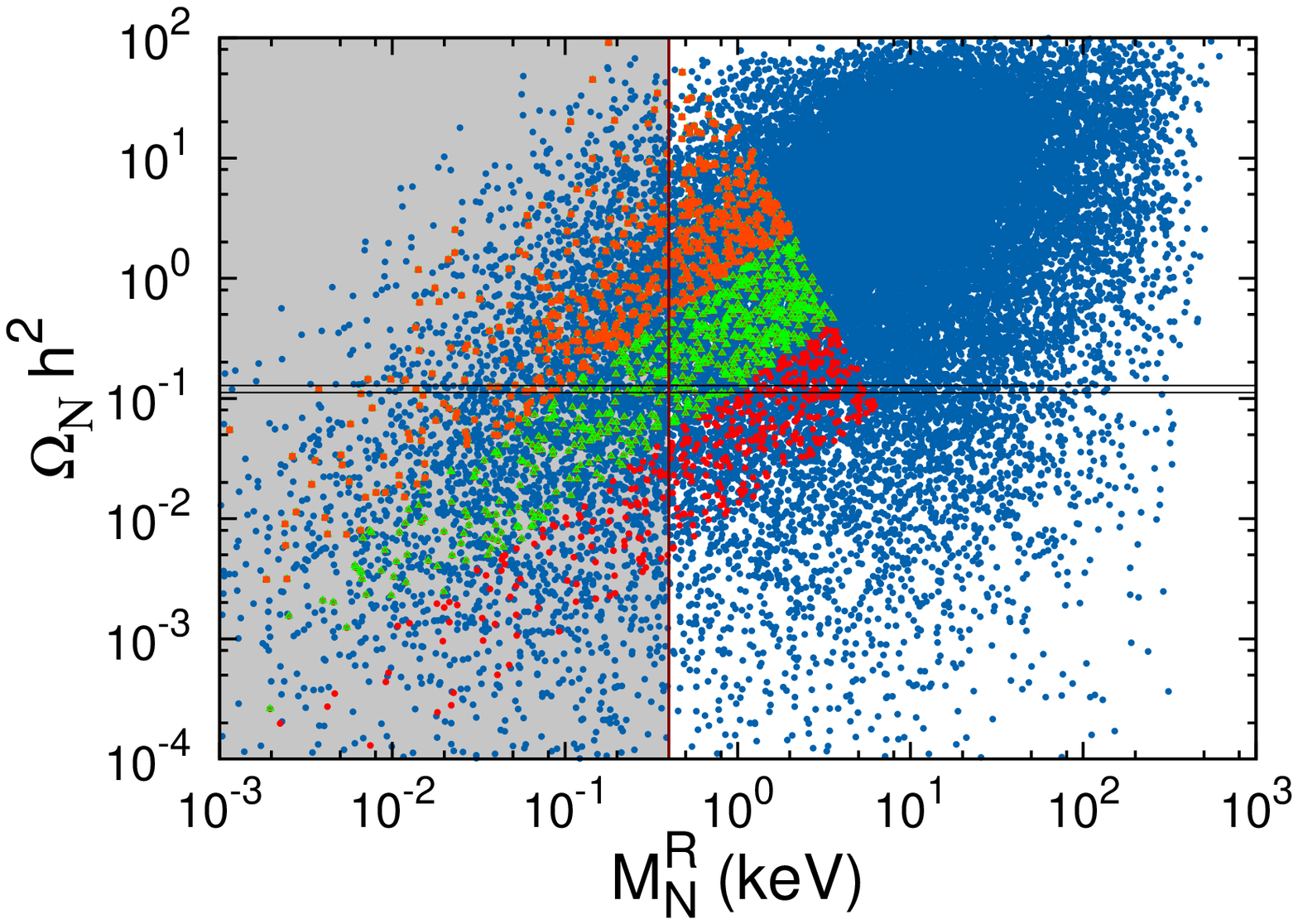}
\caption{
Scatter plot in the ($M^R_N$--$\Omega_{N}h^{2}$) plane showing the allowed 
regions, in the heavy stop scenario. The grey region describes the lower bound 
of the sterile neutrino mass, for it to become a warm dark matter candidate. 
All the scattered points satisfy the X-ray constraints. The red-circle 
scattered points show $(m_{\nu})_{\textrm Tree} > 10^{-5}$eV. 
The green(triangle) and the orange(square) points represent 
$(m_{\nu})_{\textrm Tree} > 10^{-4}, 10^{-3}$ respectively. The horizontal band is the 3$\sigma$ allowed region for the dark matter relic abundance.}
\label{sterile-mass-mixing3}
\end{figure}
In figure \ref{sterile-mass-mixing3} we showed the variation of the relic 
density of the sterile neutrino  with its mass. The blue scattered points 
respect the X-ray constraints and the Higgs boson mass within the range 
$(123-127)$ GeV. The grey region shows the parameter space disfavored by 
the Pauli exclusion principle discussed earlier. The red-circle, 
green-triangle and orange-square points represent
tree level neutrino mass greater than $10^{-5}, 10^{-4}$ and $10^{-3}$ eV 
respectively. We observe that in order to have a sterile neutrino as a warm 
dark matter candidate in our model, the neutrino mass at the tree level has to 
be very small. 
\subsection{Neutrino masses and mixing: Inverted Hierarchy}
\label{Inverted hierarchy} 
For inverted hierarchy the best-fit values of solar and the atmospheric 
neutrino mass squared differences and the three mixing angles are as follows 
\cite{Schwetz:2011qt} 
$\Delta m_{21}^{2}= 7.62\times 10^{-5}$eV$^{2}$, $|\Delta m_{31}^{2}|=
2.43\times 10^{-3}$eV$^{2}$, $\theta_{12}=34.4^{\circ}$,
$\theta_{23}=50.8^{\circ}$ and $\theta_{13}=9.1^{\circ}$, 
where $\Delta m^2_{ij} \equiv m^2_i - m^2_j$.
The neutrino mass matrix can be obtained using 
\ba
m_{\nu}=U_{PMNS} 
\left(
\begin{array}{ccc}
m_{1}&0&0\\
0&m_{2}&0\\
0&0&m_{3}
\end{array}
\right)
U_{PMNS}^{T},
\ea
where the standard PMNS matrix $U_{\rm PMNS}$, with vanishing CP violating 
phases is of the form
\ba
U_{\rm PMNS} = \left(
\begin{array}{ccc}
c_{12}c_{13}&s_{12}c_{13}&s_{13}\\
-s_{12}c_{23}-c_{12}s_{23}s_{13}&c_{12}c_{23}-s_{12}s_{23}s_{13}
&s_{23}c_{13}\\
s_{12}s_{23}-c_{12}c_{23}s_{13}&-c_{12}s_{23}-s_{12}c_{23}s_{13}
&c_{23}c_{13}
\end{array}
\right),
\ea
and $m_1$, $m_2$ and $m_3$ are the neutrino mass eigenvalues.
Since the oscillation experiments are sensitive only to the mass 
squared differences, therefore for simplicity, we can assume the lightest 
neutrino mass $m_3$ to be zero in this case. Thus we have 
$m_1^2 = |\Delta m_{31}^{2}|$ 
and $m^2_2 = \Delta m^2_{21} + m^2_1$. For example, using the central 
values of the oscillation parameters mentioned above, the three flavor neutrino mass matrix in the inverted hierarchy case comes out to be
\ba
\label{inverted-hierarchy}
m_{\nu}^{IH}=
\left(
\begin{array}{ccc}
0.049&-0.0059&-0.0052\\
-0.0059&0.0211&-0.024\\
-0.0052&-0.024&0.0311
\end{array}
\right).
\ea
The three flavor active neutrino mass matrix in our model is composed mainly of 
the one-loop radiative corrections as discussed above because the tree level 
contribution to $(m_\nu)_{11}$ is very small in order to have the correct relic density of the keV sterile neutrino dark matter. We shall now present the 
results of our numerical analysis in order to fit the three-flavor global 
neutrino data in our model in the inverted hierarchy scenario. We shall 
confine ourselves in the parameter region which will produce the correct 
value for the lightest Higgs boson mass and where the sterile right handed 
neutrino can be a good candidate for keV warm dark matter.

\begin{figure}[t]
\centering
\includegraphics[height=3.0in, width=4in]{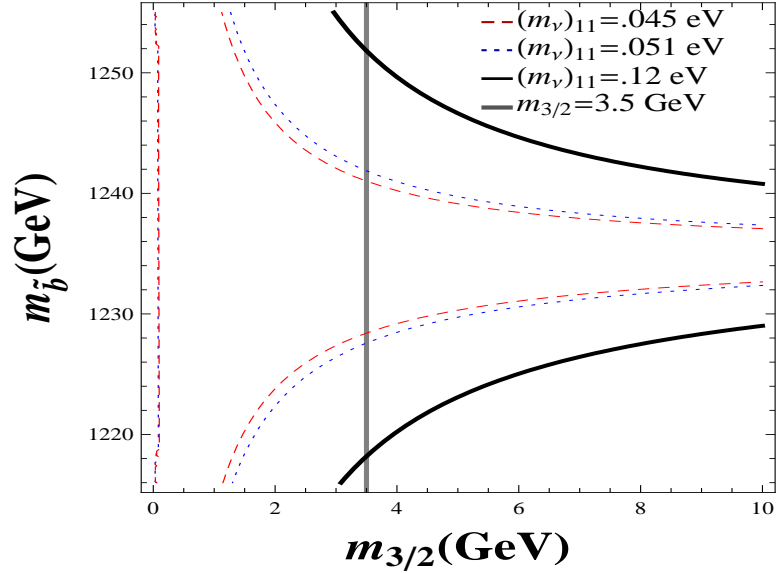}
\caption{
Contours of $(m_{\nu})_{11}$ in the ($m_{3/2}$--$m_{\tilde b}$) plane
for inverted hierarchy and $\tan\beta=10$. See text for details.
}
\label{mnu11-inv}
\end{figure}

Note that there are contributions from the tau-stau, quark-sqaurk
and neutralino-Higgs loop to the (11) element of the neutrino mass matrix 
(neglecting the tree level contribution). The trilinear R-parity violating 
couplings involved in these loop contributions are $\lambda_{133}$ and 
$\lambda_{133}^{\prime}$, which are identified with the tau and the bottom 
Yukawa couplings. The other parameters which play a crucial role in order to 
fit the (11) element of the neutrino mass matrix are $\tan\beta$, $m_{3/2}$ and $m_{\tilde b}^{2}$ (assuming that the stau-tau loop contribution 
is smaller than the other loop contributions). However, for a fixed value 
of $\tan\beta$ the trilinear couplings $\lambda_{133}$ and 
$\lambda_{133}^{\prime}$ are fixed and thus this leaves us with only two 
parameters ($m_{3/2}$ and $m_{\tilde b}^{2}$) in terms of which $(m_\nu)_{11}$ 
can be fitted. Figure \ref{mnu11-inv} presents the contour plots of 
$(m_\nu)_{11}$ in the ($m_{3/2}$--$m_{\tilde b}$) plane. Here the blue-dotted 
line corresponds to the maximum value of $(m_\nu)_{11}$ whereas 
the red-dashed line corresponds to the minimum  value of $(m_\nu)_{11}$. These 
maximum and minimum values are obtained by varying the oscillation parameters 
within the 3$\sigma$ range. Moreover, we also draw a third contour 
(the black-bold line) which represents the upper bound on $(m_{\nu})_{11}$
as obtained by the neutrinoless double beta decay experiments kamLAND-Zen 
and EXO-200 \cite{kamLAND-Zen,EXO-200}.

In order to produce this figure we fixed all the other parameters at values
corresponding to BP-4. The grey line represents a gravitino mass of 3.5 GeV. 
On the right hand side of this grey vertical line the mass of the sterile 
neutrino is $\geq$ 0.41 keV. Moreover, this plot shows the allowed range of 
sbottom mass, required to fit $(m_\nu)_{11}$ for fixed values of $\tan\beta$ 
and $m_{3/2}$. For example, we see that for $\tan\beta =10$, $m_{3/2}$ = 
3.5 GeV, the sbottom mass is allowed in the range (1228-1242) GeV. For larger 
values of $\tan\beta$ the allowed range of $m_{\tilde b}$ increases. 
\subsubsection*{Bounds on the trilinear RPV couplings for inverted hierarchy}
\label{Bounds-RPV-couplings-inverted-hierarcht}
By varying the neutrino oscillation parameters within their 
3$\sigma$ allowed ranges we can get maximum and minimum values for 
different neutrino mass matrix elements. These allowed ranges of neutrino
mass matrix elements can be translated into a lower and an upper bound 
on different trilinear R-parity violating couplings involved. With the choice
of other parameters as presented in BP-4, we present in Table 
\ref{rpv-bound-inverted} the bounds on $\lambda$ and $\lambda^\prime$ type 
couplings as functions $m_{\tilde b}$ and $m_{3/2}$ for a particular value 
of $\tan\beta$ = 10. 
\begin{table}[ht]
\centering
\caption{Bounds on $\lambda_{ijk}$ and $\lambda_{ijk}^{'}$
couplings for $\tan\beta$ = 10 and for inverted hierarchy}
\begin{tabular}{|c| c| c|}
\hline \hline
Couplings & Bounds for BP-4 & Existing bounds \\
\hline
$|\lambda_{233}^{\prime}|$ & $(2.37\times10^{-7}-1.03\times10^{-6})
\left(\frac{m_{\tilde b}}{100 ~{\rm GeV}}\right)^{2} 
\left(\frac{10 ~{\rm GeV}}{m_{3/2}}\right)$ & $6.8\times10^{-3}\cos\beta$\\
$|\lambda_{333}^{\prime}|$ & $(2.84\times10^{-7}-1.04\times10^{-6})
\left(\frac{m_{\tilde b}}{100 ~{\rm GeV}}\right)^{2}
\left(\frac{10 ~{\rm GeV}}{m_{3/2}}\right)$ & $1.305\cos\beta$\\
$|\lambda_{232}^{\prime}\lambda_{223}^{\prime}|$ & $\left(2.11-4.3\right)
\times10^{-5} \left(\frac{m_{\tilde b}}{100 ~{\rm GeV}}\right)^{2}
\left(\frac{10 ~{\rm GeV}}{m_{3/2}}\right)$ & $(2\times 10^{-3})\cos^{2}\beta
\left(\tilde\nu_{L_{2}}\tilde u_{L_{3}}\right)^{2}$\\
$|\lambda_{223}^{\prime}\lambda_{332}^{\prime}|$ & $\left(2.82-3.34\right)
\times10^{-5} \left(\frac{m_{\tilde b}}{100 ~{\rm GeV}}\right)^{2}
\left(\frac{10 ~{\rm GeV}}{m_{3/2}}\right)$ & -\\
$|\lambda_{323}^{\prime}\lambda_{332}^{\prime}|$ & $\left(2.38-4.64\right)
\times10^{-5} \left(\frac{m_{\tilde b}}{100 ~{\rm GeV}}\right)^{2}
\left(\frac{10 ~{\rm GeV}}{m_{3/2}}\right)$ & -\\
\hline
\end{tabular}
\label{rpv-bound-inverted}
\end{table}
Note that these bounds are independent of the choices of other parameters 
shown in different benchmark points because they are calculated from the 
neutrino mass matrix elements which get contributions only from the one loop 
corrections.
\subsection{Neutrino masses and mixing: Normal hierarchy}
\begin{figure}[t]
\centering
\includegraphics[height=3.0in, width=4in]{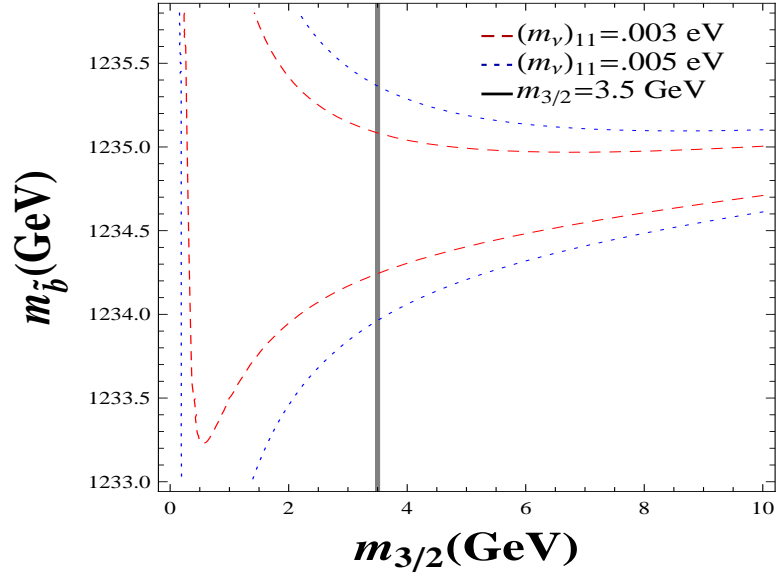}
\caption{
Contours of $(m_{\nu})_{11}$ in the ($m_{3/2}$--$m_{\tilde b}$) plane
for normal hierarchy case and $\tan\beta=10$.
\label{mnu11-nor}
}
\end{figure}

In the case of normal hierarchy best-fit values of the neutrino 
oscillation parameters are given as \cite{Schwetz:2011qt} 
$\Delta m_{21}^{2}=7.62\times
10^{-5}$ eV$^{2}$, $|\Delta m_{31}^{2}|=2.55\times 10^{-3}$ eV$^{2}$ 
and the three mixing angles are $\theta_{12} =34.4^{\circ}$, 
$\theta_{23}=51.5^{\circ}$ and $\theta_{13}=9.1^{\circ}$. With these values 
and assuming that $m_1$ = 0, $m^2_2 = \Delta m^2_{21}$ and $m^2_3 
= |\Delta m^2_{31}|$ the neutrino mass matrix in the case of normal hierarchy 
turns out to be
\ba
m_{\nu}^{NH}&=&\left(
\begin{array}{ccc}
0.0039&0.0082&0.0014\\
0.0082&0.0318&0.021\\
0.0014&0.021&0.023
\end{array}
\right).
\ea
Figure \ref{mnu11-nor} presents the 
contour of $(m_\nu)_{11}$ in the ($m_{3/2}$--$m_{\tilde b}$) plane in the case 
of normal hierarchy. Here the blue-dotted line corresponds to the maximum value 
of $(m_\nu)_{11}$ = 0.005 eV whereas the red-dashed line corresponds to the 
minimum value of $(m_\nu)_{11}$ = 0.003 eV. Once again these maximum and 
minimum values are obtained by varying the oscillation parameters within their 
3$\sigma$ range. 

The right side of the 3.5 GeV gravitino mass line can produce a keV sterile 
neutrino warm dark matter with a mass greater than 0.41 keV. The values of 
other parameters correspond to BP-4. Here we have chosen a small 
$\lambda_S = 10^{-4}$, which requires heavy stops to produce a $\sim$ 125 GeV 
Higgs boson. Looking at this figure one can also see the range of sbottom mass 
required to fit the value of $(m_\nu)_{11}$ for a fixed value of $\tan\beta$ 
and $m_{3/2}$. If we take a large value of $\lambda_{s}$, then the 
$m_{\tilde b}$ mass range changes slightly but the essential feature remains 
the same. 
\subsubsection*{Bounds on the trilinear RPV couplings for normal hierarchy}
\label{Bounds-on-RPV-couplings-for-normal-hierarchy}
One can also constrain the trilinear R-parity violating couplings in the case 
of normal hierarchy after analyzing the other elements of the neutrino mass 
matrix in the light of neutrino data. The resulting bounds are shown in 
Table \ref{rpv-bound-normal}.
\begin{table}[ht]
\centering
\caption{Bounds on $\lambda_{ijk}$ and $\lambda_{ijk}^{'}$
couplings for $\tan\beta$ = 10 and for normal hierarchy}
\begin{tabular}{|c| c| c|}
\hline \hline
Couplings & Bounds for BP-4 & Existing Constraints\\
\hline
$|\lambda_{233}^{\prime}|$ & $(8.07\times 10^{-7}-1.2\times 10^{-6})
\left(\frac{m_{\tilde b}}{100 ~{\rm GeV}}\right)^{2} 
\left(\frac{10 ~{\rm GeV}}{m_{3/2}}\right)$ & $6.8\times10^{-3}\cos\beta$\\
$|\lambda_{333}^{\prime}|$ & $(3.74\times10^{-8}-6.11\times10^{-7})
\left(\frac{m_{\tilde b}}{100 ~{\rm GeV}}\right)^{2}
\left(\frac{10 ~{\rm GeV}}{m_{3/2}}\right)$ & $1.305\cos\beta$\\
$|\lambda_{232}^{\prime}\lambda_{223}^{\prime}|$ & $\left(2.5-4.7\right)
\times10^{-5} \left(\frac{m_{\tilde b}}{100 ~{\rm GeV}}\right)^{2}
\left(\frac{10 ~{\rm GeV}}{m_{3/2}}\right)$ & $(2\times 10^{-3})\cos^{2}\beta
\left(\tilde\nu_{L_{2}}\tilde u_{L_{3}}\right)^{2}$\\
$|\lambda_{223}^{\prime}\lambda_{332}^{\prime}|$ & $\left(2.4-3.0\right)
\times10^{-5} \left(\frac{m_{\tilde b}}{100 ~{\rm GeV}}\right)^{2}
\left(\frac{10 ~{\rm GeV}}{m_{3/2}}\right)$ & -\\
$|\lambda_{323}^{\prime}\lambda_{332}^{\prime}|$ & $\left(2.5-4.69\right)
\times10^{-5} \left(\frac{m_{\tilde b}}{100 ~{\rm GeV}}\right)^{2}
\left(\frac{10 ~{\rm GeV}}{m_{3/2}}\right)$ & -\\
\hline
\end{tabular}
\label{rpv-bound-normal}
\end{table}
Bounds on trilinear R-parity violating couplings from various other studies can be found 
in \cite{Staub,Dreiner,Dreiner1,Kundu,Kundu-1, Kundu-2,Gouvea,Huitu,
Huitu:1997bi,GB,Agashe,Abada}.
\section{Case with large neutrino Yukawa coupling}
\begin{figure}[t]
\centering
\includegraphics[height=3.0in, width=4in]{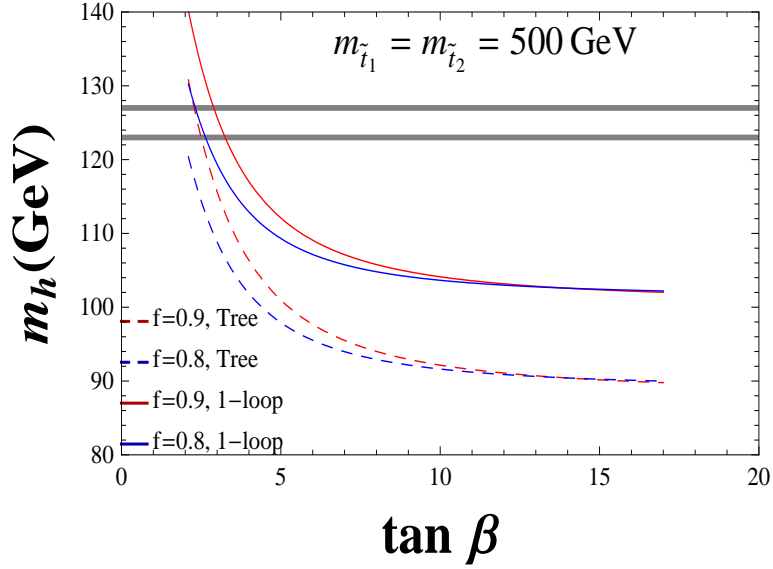}
\caption{
The variation of the lightest Higgs boson mass with $\tan\beta$.
The dashed lines represent the Higgs boson mass at the tree level 
and the continuous lines represent the Higgs boson mass after radiative 
correction is added for a stop mass of 500 GeV. Here red corresponds to 
$f = 0.9$ whereas blue corresponds to $f = 0.8$.} 
\label{tree-loop-higgs-mass}
\end{figure}

While discussing the sum rules in the scalar sector, we observed that 
the lightest Higgs boson mass receives an additional tree level contribution
due to the presence of the neutrino Yukawa term 
$f\hat H_{u}\hat L_{a}\hat N^{c}$ in the superpotential. In the minimal 
supersymmetric standard model (MSSM) one requires a very large loop 
correction in order to fit the Higgs boson mass in the range of (123 -- 127) 
GeV \cite{Hall:2011aa}. In the next-to-minimal supersymmetric model (NMSSM), 
the $\mu$ term is dynamically generated through a $\lambda_{S} SH_{u}H_{d}$ 
term in the superpotential and the tree level Higgs boson mass receives a 
correction proportional to $\lambda_{S}^{2}$ \cite{Hall:2011aa}. Similarly in 
the singlet-triplet extension of the MSSM, a tree level correction to the Higgs boson mass proportional to $\lambda_{S}^{2}$ and $\lambda_{T}^{2}$ is 
obtained \cite{Mohanty}.
However, in this model these tree level contributions to the lightest Higgs 
boson mass are absent but because of the presence of the neutrino Yukawa 
coupling $f$ an additional contribution 
$(\Delta m^2_h)_{\rm tree} = f^2 v^2 \sin^2 2\beta$ is obtained. 
In figure \ref{tree-loop-higgs-mass} we show the variation of the lightest 
Higgs boson mass in this model as a function of $\tan\beta$.
We can observe from this figure that for a low value of $\tan\beta$ 
the Higgs boson mass of $\sim 125 ~{\rm GeV}$ can be achieved with $f=0.9$, 
even at the tree level. Moreover, we find that $m_{{\tilde t}_1} 
= m_{{\tilde t}_2} = 500 ~{\rm GeV}$ is sufficient enough to provide 
the correct Higgs boson mass through radiative corrections for a slightly 
larger value of $\tan\beta$.

In this case the parameter $M_{R}$, included in the superpotential 
as $M_{R}N^{c}S$, is very large and thus the sterile neutrino becomes very 
heavy (see eq. (\ref{relation-MR-f})). In such a situation the lightest 
neutralino with a large bino component becomes the LSP with a very small 
mass of a few hundred MeV. The mass of the LSP is essentially controlled 
by the R-symmetry violating Majorana gaugino mass parameter $M_{1}$. We 
show the benchmark points corresponding to the large $f$ scenario in 
Table \ref{Table-6},
\begin{table}[ht]
\centering
\begin{tabular}{|c| c|}
\hline \hline
Parameters & BP-7 \\
\hline
$M_{1}^{D}$ & 800 GeV\\
$M_{2}^{D}$ & 580 GeV\\
$\tan\beta$ & 2.6\\
$\lambda_{S}$& $10^{-5}$\\
$\lambda_{T}$& $\lambda_{S}\tan\theta_{W}\sim$ $5.5\times 10^{-6}$\\
$\mu$ & 200 GeV\\
$t_S$ & $(200)^{3}$\\
b$\mu_{L}$ & $-(200)^{2}$ (GeV)$^{2}$\\
$m_{S}$ & 7.39 TeV\\
$m_{T}$ & 7.7 TeV\\
$v_{S}$ & 0.5 GeV \\
$v_{T}$ & 0.1 GeV \\
f & 0.9\\
$M_{R}$ &  7.4 TeV\\
$m_{\tilde t_{1}}=m_{\tilde t_{2}}$ & 500 GeV\\
$b_{S}$ & 1 TeV  \\
$b_{T}$ & 1 TeV  \\
$m_{3/2}$ & 20 GeV \\
\hline
$m_{h}$ & 125.5 GeV \\
$(m_{\nu})_{\textrm{Tree}}$ & 0.049 eV \\
$m_{\tilde \chi}^{0}$ & 167 MeV\\
\hline
\end{tabular}
\caption{A benchmark point with large f and small $\lambda_S$ and 
$\lambda_{T}$} 
\label{Table-6}
\end{table}
where a tree level neutrino mass of $0.049 ~{\rm eV}$ and a lightest 
neutralino mass of 167 MeV are obtained. Several studies can be found 
in the literature \cite{Kittel, Ota, Kim, Sarkar, Choudhury} concerning very 
light neutralinos. These include non universal gaugino mass models and R-parity violation. In the context of NMSSM with R-parity conservation, a few hundred 
MeV bino-like lightest neutralino has been studied as a dark matter 
candidate and it has been shown that one can avoid the overproduction of such 
a light neutralino in the early universe through efficient annihilations 
\cite{Gunion:2005rw}. 

However, in our case the MeV neutralino LSP can decay through R-parity 
violating channels. Note, that in this case the gravitino with a mass 
of $\sim$ 10 GeV decays mainly to the lightest neutralino + photon final 
state and has a lifetime of $\sim 10^{12} ~{\rm sec}$. Such a gravitino 
will decay after the big-bang nucleosynthesis (BBN) producing an unacceptable
amount of entropy. This conflicts with the predictions of BBN if one 
assumes the standard big-bang cosmology and results in a constraint on the 
gravitino mass to be $m_{3/2} > ~10 ~{\rm TeV}$ \cite{weinberg}. However, 
this constraint on the gravitino mass can be avoided if one assumes that the 
universe had gone through an inflationary phase and in order to
avoid the strong constraints obtained from the photo-dissociation of the 
light elements because of the radiative decay of the gravitino, one arrives 
at the upper bound on the reheating temperature of the universe 
$T_R \lsim 10^6$ GeV \cite{Sarkar, moroi}. In this case the gravitino is 
a stable particle in the collider time scale. However, it cannot be a candidate for dark matter because of its small lifetime in the cosmological time scale. 
Implications of such a scenario in the context of collider studies and dark 
matter requires further investigations and we shall postpone this for a 
future work. 
\section{Conclusions and Outlook}
We have studied a supersymmetric model of neutrino masses and mixing with 
an $U(1)_R$ symmetry and a single right handed neutrino superfield. In this 
model the R-symmetry is identified with lepton number in such a way that the 
lepton numbers of the standard model fermions are the same as their R-charges 
but with a negative sign. The neutral gauginos are Dirac fermions in this model and one needs to introduce additional chiral superfields in the
adjoint representations of the gauge groups. The right-handed neutrino with an 
appropriate R-charge allows one to write down neutrino Yukawa interactions 
respecting the $U(1)_R$ symmetry. After the electroweak symmetry breaking one 
of the sneutrinos (we choose it to be the electron sneutrino) develops a 
non-zero vacuum expectation value, which can be significant because it is not 
constrained by small neutrino masses. In the neutral fermion sector we have 
mixing among the neutralinos, the electron-neutrino and the right handed 
neutrino consistent with the R-symmetry and that results in 
a small Dirac neutrino mass at the tree level. The scalar sector of this 
model can accommodate a Higgs boson with $\sim$ 125 GeV mass. This can be 
achieved even at the tree level with the help of a large Dirac neutrino 
Yukawa coupling ($f \sim 1$) or including the one loop radiative corrections 
to the tree level mass of the Higgs boson. A very important property of this 
R-symmetric model is the existence of a subset of R-parity violating 
interactions in the superpotential parametrized by $\lambda$ and 
$\lambda^\prime$ in the literature. 

There are two massless active neutrinos at the tree level, which acquire 
non-zero masses through one-loop radiative corrections when small R-symmetry 
breaking effects are turned on through a small gravitino mass. In this work 
we confine ourselves in a situation where the breaking of R-symmetry is 
communicated to the visible sector through anomaly mediated supersymmetry 
breaking. This results in small Majorana gaugino masses as well as trilinear 
scalar couplings, which were zero in the R-conserving limit. 
Depending on the size of the R-symmetry breaking order parameter 
(gravitino mass $m_{3/2}$ in this case), one can either generate 
a pair of almost degenerate neutrinos forming a pseudo-Dirac neutrino or 
two distinct light Majorana neutrinos from the neutralino-neutrino mass 
matrix at the tree level. Our analysis shows that none of these situations 
can accommodate the results from LSND experiments with a possible neutrino mass eigenstate at $\sim$ 1.2 eV. On the other hand, there exists a possibility of 
having a Majorana sterile neutrino with a mass of the order of a few keV, which can be a good candidate for warm dark matter. A detail scan of our parameter 
space shows that there are allowed regions where the constraints on this keV 
dark matter coming from X-ray observations can be satisfied and this keV 
sterile neutrino can account for the dark matter relic density measured at 
the PLANCK and WMAP experiments. At the same time there exists an active 
neutrino acquiring a very small mass at the tree level. All these allowed 
points in the parameter space are consistent with a $\sim$ 125 GeV light Higgs 
boson. We have also identified two distinct cases of heavy and light stop 
masses consistent with the Higgs boson mass, dark matter relic density 
constraint and a small tree level mass of the neutrino. The collider 
signatures of these two cases should be explored further which can possibly 
provide some testable predictions at the LHC. Because of the mixing in the 
neutralino, chargino and the scalar mass matrices the sparticles can have 
novel decay modes leading to interesting final states in $pp$ collision and can be studied in a similar way as presented in Ref.\cite{Kumar}.

We investigate the light active neutrino sector and try to fit the three flavor global neutrino data by incorporating one loop radiative corrections to the 
(3 $\times$ 3) light neutrino mass matrix. We choose certain benchmark points 
for our numerical analysis and show that one can obtain bounds on the trilinear R-parity violating couplings in the superpotential from neutrino data as a 
function of the R-symmetry breaking order parameter ($m_{3/2}$). We further pay a special attention to the situation with a large Dirac neutrino Yukawa 
coupling $f$ and demonstrate that a large $f$ can induce additional tree level
contribution to the lightest Higgs boson mass to be consistent with the Higgs 
boson mass measurement at the LHC experiments. Even with such a large value 
of $f$, a small Majorana mass for the light active neutrino can be generated at the tree level. A very interesting feature of this scenario is the existence of a few hundred MeV lightest neutralino LSP with a substantial bino component. In this R-parity violating scenario, this MeV lightest neutralino LSP can decay 
into final states involving standard model fermions and can avoid the
constraints on such a light MeV neutralino from its overproduction in the early universe. The gravitino is the NLSP in this case with a mass $m_{3/2} \sim$ 
10 GeV and it is a stable particle in the collider time scale with a lifetime 
of $\sim 10^{12} ~{\rm sec}$. This is cosmologically consistent as long as the 
reheating temperature $T_R \lsim 10^6$ GeV. It would be really interesting to 
see the phenomenological and cosmological implications of this MeV neutralino 
scenario in detail, and in particular, at the LHC. However, such a dedicated 
analysis is beyond the scope of this paper.    
\section{Acknowledgments}
SC would like to thank the Council of Scientific and Industrial Research, 
Govt. of India for the financial support received as a Senior Research Fellow. 
He also thanks K. Huitu, J. Maalampi, K. Kainulainen, P. Bandyopadhyay, 
S. Di Chiara and Subhadeep Mondal for many useful comments and suggestions.

\end{document}